\documentclass[aps,prb,showpacs,twocolumn,floats]{revtex4}
\usepackage{amssymb}

\usepackage{graphicx,psfig}
\usepackage{dcolumn}
\usepackage{bm}

\input{tcilatex}

\begin{document}

\title{Many-body Landau-Zener effect at fast sweep}
\author{D. A. Garanin and R. Schilling}
\affiliation{Institut f\"{u}r Physik, Johannes
Gutenberg-Universit\"{a}t, D-55099 Mainz, Germany}
\date{22 December 2004}

\begin{abstract}
The asymptotic staying probability $P$ in the Landau-Zener effect with
interaction is analytically investigated at fast sweep, $\varepsilon =\pi
\Delta ^{2}/(2\hbar v)\ll 1$. We have rigorously calculated the value of $%
I_{0}$ in the expansion $P\cong 1-\varepsilon +\varepsilon
^{2}/2+\varepsilon ^{2}I_{0}$ for arbitrary couplings and relative resonance
shifts of individual tunneling particles. The results essentially differ
from those of the mean-field approximation. It is shown that strong
long-range interactions such as dipole-dipole interaction (DDI) generate
huge values of $I_{0}$ because flip of one particle strongly influences many
others. However, in the presence of strong static disorder making resonance
for individual particles shifted with respect to each other the influence of
interactions is strongly reduced. In molecular magnets the main source of
static disorder is the coupling to nuclear spins. Our calculations using the
actual shape of the Fe$_{8}$ crystal studied in the the Landau-Zener
experiments [Wernsdorfer et al, Europhys. Lett. \textbf{50}, 552 (2000)]
yield $I_{0}\ $that is in a good agreement with the value extracted from the
experimental data.
\end{abstract}
\pacs{ 03.65.-w, 75.10.Jm}

\maketitle

%\pacs{ 03.65.-w, 75.10.Jm}

\section{Introduction}

\begin{figure}[t]
\unitlength1cm
\begin{picture}(11,6)
\centerline{\psfig{file=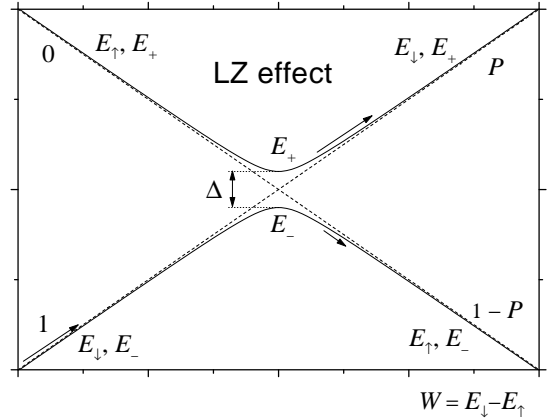,angle=-90,width=9cm}}
\end{picture}
\caption{A pair of tunnel-splitted levels vs. energy bias $W(t)$. Here $%
E_{\downarrow }$ and $E_{\uparrow }$ are the bare energy levels ($\Delta =0$%
), whereas $E_{\pm }$ are the exact adiabatic energy levels of Eq.\ (\ref
{EpmDef}). $P$ denotes the probability to remain in the (bare) state $\left|
\downarrow \right\rangle $ after crossing the resonance. }
\label{LZEffect}
\end{figure}

Landau-Zener (LZ) effect \cite{lan32,zen32} (see also Refs.\
\onlinecite
{stu32,akusch92,dobzve97}) is a well known quantum phenomenon of transitions
at avoided level crossing, see Fig.\ \ref{LZEffect}. LZ effect was
encountered mainly in physics of atomic and molecular collisions (see, e.g.,
Refs.\ \onlinecite{chi74,nak02} and references therein). In the
time-dependent formulation, the LZ effect can be modeled by a two-level
system (TLS)
\begin{equation}
\hat{H}=-\frac{1}{2}W(t)\sigma _{z}+\frac{1}{2}\Delta \sigma _{x},
\label{HamLZ1Part}
\end{equation}
where $\sigma _{\alpha },$ $\alpha =x,y,z$ are the Pauli matrices and
\begin{equation}
W(t)\equiv E_{\downarrow }(t)-E_{\uparrow }(t)  \label{WtDef}
\end{equation}
is the time-dependent bias of the two bare ($\Delta =0)$ energy levels. The
general state of this model and the probability $P(t)$ to stay in the $%
\left| \downarrow \right\rangle $ state can be written as
\begin{equation}
\Psi (t)=a_{\downarrow }(t)\left| \downarrow \right\rangle +a_{\uparrow
}(t)\left| \uparrow \right\rangle ,\qquad P(t)=\left| a_{\downarrow
}(t)\right| ^{2}.  \label{PsiPDef}
\end{equation}
The initial condition is $W(-\infty )=-\infty $ and $P(-\infty )=1.$ If $%
W(t) $ changes fast, the system does not have enough time for transition to
the state $\left| \uparrow \right\rangle $ and it practically remains in the
state $\left| \downarrow \right\rangle ,$ thus $P(t)\cong 1.$ In the
opposite case of slow $W(t)$ the system mainly remains at the lower of the
adiabatic energy levels
\begin{equation}
E_{\pm }(t)=\pm \frac{1}{2}\sqrt{W^{2}(t)+\Delta ^{2}},  \label{EpmDef}
\end{equation}
thus the asymptotic staying probability $P(\infty )\cong 0.$

The time-dependent Schr\"{o}dinger equation for the Hamiltonian of Eq.\ (\ref
{HamLZ1Part}) can be solved exactly for the linear sweep $W(t)=vt$, the
result for the asymptotic staying probability being
\begin{equation}
P\equiv P(\infty )=e^{-\varepsilon },\qquad \varepsilon \equiv \frac{\pi
\Delta ^{2}}{2\hbar v}.  \label{PLZDef}
\end{equation}
As the Schr\"{o}dinger equation (SE) for a spin 1/2 is mathematically
equivalent to the classical dissipationless Landau-Lifshitz equation (LLE),
the effect can be viewed upon as a rotation of a classical magnetization
vector. \cite{chugar02,gar03prb}

\begin{figure}[t]
\unitlength1cm
\begin{picture}(11,6)
\centerline{\psfig{file=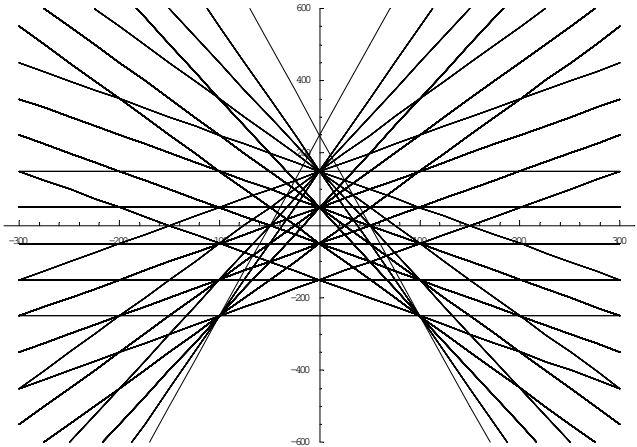,angle=-90,width=9cm}}
\end{picture}
\caption{Bare energy levels vs the sweep field $W$ for a ring of 10
two-level systems with $V_{i}=0,$ coupled by an antiferromagnetic
nearest-neighbor interaction. All the lines except the two ones with the
highest slopes that correspond to the states with all spins up and all spins
down are strongly degenerate. The energy levels for the ferromagnetic
coupling can be obtained by the upside-down transformation of this graph.}
\label{LZEffect-ManyBody}
\end{figure}

Recently the LZ effect was observed on crystals of molecular magnets Fe$_{8}$
in Refs.\ \onlinecite{werses99science,weretal00epl} (see Refs.\
\onlinecite
{wer01thesis,sesgat03}\ for a recent review). This posed a new problem of
the many-body LZ effect that can be described by the Hamiltonian of the
transverse Ising model
\begin{equation}
\hat{H}=-\frac{1}{2}\sum_{i}\left\{ \left[ W(t)-V_{i}\right] \sigma
_{iz}+\Delta \sigma _{ix}\right\} -\frac{1}{2}\sum_{ij}J_{ij}\sigma
_{iz}\sigma _{jz},  \label{Ham}
\end{equation}
where $V_{i}$ are the local shifts of the resonances for individual TLSs
that can be induced, for instance, by nuclear spins\cite{weretal00prl} and $%
J_{ij}$ is their coupling. The SE for such a system of $N\gg 1$ coupled TLSs
contains $2^{N}$ time-dependent coefficients of the wave function. In the
language of the time-dependent bare energy levels, there are $2^{N}$
different lines that cross each other at different values of $W,$ see Fig.\
\ref{LZEffect-ManyBody}.

Eq.\ (\ref{Ham}) can be extended by taking into account coupling to
environmental degrees of freedom such as phonons, as was done for
one-particle LZ effect in Refs.\ \onlinecite{aoram91prb,kaynak98prb}. One
can also consider dynamics of nuclear spins\cite{sinpro03prb} that are
treated in a simplified way as static disorder in Eq.\ (\ref{Ham}). As
quantum dynamics of many-body systems far from equilibrium and coupled to
environment is very involved, simplified theories using postulated rate
equations (neglecting quantum-mechanical coherence) have been proposed for
both the LZ effect and the relaxation out of a prepared state. \cite
{prosta98prl,cucforretadavil99epjb,alofer01prl,liuetal02prb,feralo03prl} All
these theories consider the molecular field at the cite $i$ to determine
whether the particle $i$ is in the vicinity of the resonance and thus can
flip. This means that these theories are based on the mean-field
approximation (MFA).

One can perform the MFA for the dissipationless system described by Eq.\ (%
\ref{Ham}) by considering a single particle undergoing a LZ transition in
the effective field being the sum of the externally sweeped field $W(t)$ and
the molecular field from other particles that is determined self
consistently. \cite{hamraemiysai00} This is a model of the \emph{nonlinear}
LZ effect that was applied to tunneling of the Bose-Einstein condensate.
\cite{zobgar00,wuniu00} Again the problem can be reformulated in terms of a
classical \emph{nonlinear} LLE, see Ref.\ \onlinecite{gar03prb}, although
the LZ effect is a quantum-mechanical phenomenon! The MFA solution shows
that ferromagnetic interactions suppress transitions since in this case the
total effective field changes faster than $W(t)$ and thus the effective
value of the sweep-rate parameter $\varepsilon $ in Eq.\ (\ref{PLZDef})
becomes smaller, while antiferromagnetic interactions enhance transitions.
There are, however, models with couplings of different signs for which
contributions to the molecular fields cancel each other and the MFA wrongly
predicts no effect of interactions.

An important special case of Eq.\ (\ref{Ham}) is the so-called ``spin-bag''
model with the same coupling $J$ for any two spins. \cite
{hamraemiysai00,gar03prb} In the limit $N\rightarrow \infty $ with $%
J_{0}\equiv (N-1)J=\mathrm{const}$ the MFA for the spin-bag model becomes
exact. \ This model can be exactly mapped onto the model with the spin $%
S=N/2 $ and the Hamiltonian $\widehat{H}=-W(t)S_{z}-\Delta S_{x}-2JS_{z}^{2}.
$ In contrast to the model with a general interaction that is difficult to
solve numerically, the spin-bag model can be numerically solved up to pretty
big values of $N$ that allows to check approaching to the mean-field limit $%
N\rightarrow \infty $ for the original model and to the classical limit for
the equivalent spin-$S$ model. \cite{gar03prb} A surprising result of Ref.\ %
\onlinecite{gar03prb} that also should be valid for a general interaction is
that MFA becomes exact in the linear order in small $J,$ even without going
to the limit $N\rightarrow \infty .$ Quantum corrections to the dynamics of
the spin-$S$ model in the limit $S\gg 1$ (i.e., deviations from the
mean-field results for the original spin-bag model) were systematically
investigated in Ref.\ \onlinecite{garsch04prb}.

For models with realistic finite-range interactions that are not small, the
applicability of the MFA to the description of the LZ effect is not
justified. In particular, one can expect large deviations from the MFA for
competing interactions such as the dipole-dipole interaction (DDI).
Fortunately, in the fast-sweep limit $\varepsilon \ll 1$ one can construct a
rigorous perturbative expansion of the staying probability $P$ for Eq.\ (\ref
{Ham}) in powers of $\Delta ^{2}.$ The result can be written in the form\cite
{garsch03eprint}
\begin{equation}
P\cong 1-\varepsilon +\varepsilon ^{2}/2+\varepsilon ^{2}I_{0}\cong
e^{-\varepsilon }+\varepsilon ^{2}I_{0}  \label{PExpansion}
\end{equation}
with $I_{0}$ depending on the coupling $J_{ij},$ resonance shifts $V_{i}$
and the sweep rate $v.$ The applicability of this expansion requires $%
\varepsilon I_{0}\lesssim 1.$ Eq.\ (\ref{PExpansion}) shows whether
interactions suppress transitions ($I_{0}>0$) or enhance them ($I_{0}<0$)
that is not clear in the case of the DDI. On the other hand, this rigorous
result can be used to test the applicability of the MFA and other
approximations that can be suggested to describe the many-body LZ effect.
The results of experiments measuring $P$ can be parametrized with the help
of the effective splitting $\Delta ^{\mathrm{eff}}$ calculated from Eq.\ (%
\ref{PLZDef}):
\begin{equation}
\Delta ^{\mathrm{eff}}=\sqrt{\frac{2\hbar v}{\pi }\ln \frac{1}{P}},
\label{DeltaeffDef}
\end{equation}
as was done in Ref.\ \onlinecite{weretal00epl}. In the fast-sweep limit $%
\Delta ^{\mathrm{eff}}$ should have the form following from Eq.\ (\ref
{PExpansion}):
\begin{equation}
\Delta ^{\mathrm{eff}}=\Delta \sqrt{\frac{1}{\varepsilon }\ln \frac{1}{P}}%
\cong \Delta \left( 1-\frac{1}{2}I_{0}\varepsilon \right) .
\label{DeltaeffExpansion}
\end{equation}
Thus if $I_{0}\approx \mathrm{const},$ as is the case for strong couplings,
\cite{garsch03eprint} Eq.\ (\ref{DeltaeffExpansion}) allows to determine $%
\Delta $ and $I_{0}$ from the linear extrapolation to $\varepsilon
\rightarrow 0,$ see Fig.\ \ref{Fig-Fe8-Wernsdorfer} below.

The aim of this article is to explain the derivation of Eq.\ (\ref
{PExpansion}) in a more detail, provide a comparison with the MFA result,
and investigate the role of the DDI in crystals of molecular magnets of
non-ellipsoidal shape where the magnetostatic field is inhomogeneous and
thus the resonance fields for molecules in different parts of the crystal
are different. The latter is needed to make a comparison with the results of
Ref.\ \onlinecite{weretal00epl} on a crystal of a rectangular shape.

The remainder of this paper is organized as follows. In Sec. \ref{Sec-PT} we
construct the perturbation scheme for the many-body LZ effect at fast sweep
and derive the general expression for $I_{0}.$ In Sec. \ref{Sec-Analysis} we
analyze different limiting forms of $I_{0}$ that will be used below. In this
section we also consider the role of static disorder, mainly due to nuclear
spins, described by $V_{i}$ in Eq.\ (\ref{Ham}) and perform averaging over
Gaussian distribution of $V_{i}.$ In Sec. \ref{Sec-DDI} we consider the
influence of the DDI on the Landau-Zener transitions in samples of the
ellipsoidal shape, where the dipolar field is homogeneous, as well in
samples of a general shape in the case of strong static disorder. We compare
our results with the experimental data of Ref.\ \onlinecite{weretal00epl} in
Sec. \ref{Sec-experiment}. In Sec. \ref{Sec-discussion} we recollect our
main results and make some proposals for future research.

\section{Perturbation theory for fast sweep}

\label{Sec-PT}

We consider the transverse-field Ising model, Eq.\ (\ref{Ham}), with the
time-linear sweep $W(t)=vt$. The wave function of the system of $N$
tunneling particles (TLSs) can be written as the expansion over the
direct-product states
\begin{eqnarray}
\Psi (t) &=&\sum_{m_{1},\ldots ,m_{N}=-1,1}\mathcal{C}_{m_{1},\ldots
,m_{N}}(t)\psi _{m_{1}}\otimes \ldots \otimes \psi _{m_{N}}  \nonumber \\
\psi _{-1} &=&\left(
\begin{array}{c}
0 \\
1
\end{array}
\right) =|\downarrow \rangle ,\qquad \psi _{1}=\left(
\begin{array}{c}
1 \\
0
\end{array}
\right) =|\uparrow \rangle .  \label{PsiDef}
\end{eqnarray}
The initial condition for $\Psi (t)$ is $\left| \mathcal{C}_{-1,\ldots
,-1}(-\infty )\right| =1$ whereas all other coefficients are zero, i.e., the
system starts in the state with all spins down. One can denote this state as
$\left| \downarrow \downarrow \ldots \downarrow \right\rangle .$ As the time
evolves, the state of the system becomes a superposition of all possible
basis states in Eq.\ (\ref{PsiDef}). One can write the wave function in the
form
\begin{eqnarray}
\Psi (t) &=&c_{0}(t)\left| \downarrow \downarrow \ldots \downarrow
\right\rangle +\sum_{i}c_{i}(t)\sigma _{i+}\left| \downarrow \downarrow
\ldots \downarrow \right\rangle  \nonumber \\
&&+\frac{1}{2!}\sum_{ij}c_{ij}(t)\sigma _{i+}\sigma _{j+}\left| \downarrow
\downarrow \ldots \downarrow \right\rangle +\ldots  \label{PsiForm}
\end{eqnarray}
including all possible numbers of flipped spins. Then the initial condition
becomes $\left| c_{0}(-\infty )\right| =1$ and $c_{ij}(-\infty )=0,$ etc.
The normalization condition for this wave function is
\begin{equation}
1=\left| c_{0}\right| ^{2}+\sum_{i}\left| c_{i}\right| ^{2}+\frac{1}{2!}%
\sum_{ij}\left| c_{ij}\right| ^{2}+\ldots  \label{NormPsi1}
\end{equation}
The probability for a particle at the site $i$ to stay in the initial state
is given by
\begin{eqnarray}
P_{i} &=&\left| c_{0}\right| ^{2}+\sum_{j\neq i}\left| c_{j}\right| ^{2}+%
\frac{1}{2!}\sum_{j\neq i,k\neq i}\left| c_{jk}\right| ^{2}+\ldots  \nonumber
\\
&=&1-\left| c_{i}\right| ^{2}-\sum_{j}\left| c_{ij}\right| ^{2}+\ldots
\label{PiDef}
\end{eqnarray}
where we have used Eq.\ (\ref{NormPsi1}) to simplify the expression. The
staying probability averaged over the system is
\begin{equation}
P=\frac{1}{N}\sum_{i}P_{i}=1-\frac{1}{N}\sum_{i}\left| c_{i}\right| ^{2}-%
\frac{1}{N}\sum_{ij}\left| c_{ij}\right| ^{2}-\ldots  \label{PAvrDef}
\end{equation}
The Schr\"{o}dinger equation for the coefficients in Eq.\ (\ref{PsiForm})
reads
\begin{eqnarray}
i\hbar \dot{c}_{0} &=&0\times c_{0}-\frac{\Delta }{2}\sum_{i}c_{i}  \nonumber
\\
i\hbar \dot{c}_{i} &=&E_{i}(t)c_{i}-\frac{\Delta }{2}c_{0}-\frac{\Delta }{2}%
\sum_{j}c_{ij}  \nonumber \\
i\hbar \dot{c}_{ij} &=&E_{ij}(t)c_{ij}-\frac{\Delta }{2}\left(
c_{i}+c_{j}\right) -\frac{\Delta }{2}\sum_{l}c_{ijl},  \label{SchrEq}
\end{eqnarray}
etc. Here $E_{i}$ and $E_{ij}$ are the eigenvalues of the Hamiltonian with $%
\Delta =0$ and the ground-state energy $E_{0}(t)$ subtracted
\begin{eqnarray}
E_{i}(t) &=&-W(t)+\tilde{V}_{i},\qquad  \nonumber \\
E_{ij}(t) &=&-2W(t)+\tilde{V}_{i}+\tilde{V}_{j}-4J_{ij},  \label{EDef}
\end{eqnarray}
where
\begin{equation}
\tilde{V}_{i}\equiv V_{i}+V_{i}^{(\mathrm{int})},\qquad V_{i}^{(\mathrm{int}%
)}\equiv 2\sum_{j}J_{ij}.  \label{VIntDef}
\end{equation}
Whereas $V_{i}$ that enters Eq.\ (\ref{Ham}) can be random, there is another
contribution into resonance shifts, $V_{i}^{(\mathrm{int})},$ that can
gradually change across the sample for long-range interactions such as the
DDI.

For fast sweep there is little time for spin flipping and the system remains
near the initial state: $\left| c_{0}(t)\right| \cong 1$ while all other
coefficients are small. Since spin flipping is caused by $\Delta ,$ one can
consider $\Delta $ as a formal small parameter and obtain the solution of
Eq.\ (\ref{SchrEq}) in the fast-sweep limit iteratively in powers of $\Delta
$. To this end, it is convenient to introduce slow amplitudes $\tilde{c}(t)$
according to
\begin{equation}
c_{\nu }(t)=\tilde{c}_{\nu }(t)e^{-i\Phi _{\nu }(t)},\qquad \Phi _{\nu
}(t)\equiv \frac{1}{\hbar }\int_{0}^{t}dt^{\prime }E_{\nu }(t^{\prime })
\label{PhiDef}
\end{equation}
with $\nu =i,ij$ and rewrite Eq.\ (\ref{SchrEq}) in the form
\begin{eqnarray}
\frac{d}{dt}\tilde{c}_{0} &=&\frac{i\Delta }{2\hbar }\sum_{i}\tilde{c}%
_{i}e^{-i\Phi _{i}(t)}  \nonumber \\
\frac{d}{dt}\tilde{c}_{i} &=&\frac{i\Delta }{2\hbar }\tilde{c}_{0}e^{i\Phi
_{i}(t)}+\frac{i\Delta }{2\hbar }\sum_{j}\tilde{c}_{ij}e^{i\left[ \Phi
_{i}(t)-\Phi _{ij}(t)\right] }  \nonumber \\
\frac{d}{dt}\tilde{c}_{ij} &=&\frac{i\Delta }{2\hbar }\tilde{c}_{i}e^{i\left[
-\Phi _{i}(t)+\Phi _{ij}(t)\right] }+\left( i\rightleftharpoons j\right) .
\label{SchrEqSlow}
\end{eqnarray}
In the last equation we have dropped the term with $\tilde{c}_{ijl}$ since
we are going to calculate $P$ up to $\Delta ^{4}$ and it can be shown that
at this order $\tilde{c}_{ijl}$ is irrelevant. Iterating Eqs.\ (\ref
{SchrEqSlow}) yields
\begin{eqnarray}
\tilde{c}_{0} &=&1+\left( \frac{i\Delta }{2\hbar }\right) ^{2}\tilde{c}%
_{0}^{(2)}+\ldots  \nonumber \\
\tilde{c}_{i} &=&\frac{i\Delta }{2\hbar }\tilde{c}_{i}^{(1)}+\left( \frac{%
i\Delta }{2\hbar }\right) ^{3}\tilde{c}_{i}^{(3)}+\ldots  \nonumber \\
\tilde{c}_{ij} &=&\left( \frac{i\Delta }{2\hbar }\right) ^{2}\tilde{c}%
_{ij}^{(2)}+\ldots  \label{cExpansion}
\end{eqnarray}
where we have written down only relevant terms that are given by
\begin{eqnarray}
\tilde{c}_{i}^{(1)}(t) &=&\int_{-\infty }^{t}dt^{\prime }e^{i\Phi
_{i}(t^{\prime })}  \nonumber \\
\tilde{c}_{ij}^{(2)}(t) &=&\int_{-\infty }^{t}dt^{\prime }e^{i\left[ -\Phi
_{i}(t^{\prime })+\Phi _{ij}(t^{\prime })\right] }\tilde{c}%
_{i}^{(1)}(t^{\prime })+\left( i\rightleftharpoons j\right)  \nonumber \\
\tilde{c}_{0}^{(2)}(t) &=&\int_{-\infty }^{t}dt^{\prime }\sum_{i}\tilde{c}%
_{i}^{(1)}(t^{\prime })e^{-i\Phi _{i}(t^{\prime })}  \nonumber \\
\tilde{c}_{i}^{(3)}(t) &=&\int_{-\infty }^{t}dt^{\prime }e^{i\Phi
_{i}(t^{\prime })}\tilde{c}_{0}^{(2)}(t^{\prime })  \nonumber \\
&&+\sum_{j}\int_{-\infty }^{t}dt^{\prime }\tilde{c}_{ij}^{(2)}(t^{\prime
})e^{i\left[ \Phi _{i}(t^{\prime })-\Phi _{ij}(t^{\prime })\right] }.
\label{cIter}
\end{eqnarray}
Now the expansion of $P_{i}$ of Eq.\ (\ref{PiDef}) up to $\Delta ^{4}$ has
the form
\begin{eqnarray}
P_{i}(t) &=&1-\left( \frac{\Delta }{2\hbar }\right) ^{2}\left| \tilde{c}%
_{i}^{(1)}\right| ^{2}-\left( \frac{\Delta }{2\hbar }\right) ^{4}  \nonumber
\\
&&\quad \times \left[ \sum_{j}\left| \tilde{c}_{ij}^{(2)}\right| ^{2}+2\func{%
Re}\left( \tilde{c}_{i}^{(1)\ast }\tilde{c}_{i}^{(3)}\right) \right] +\ldots
\label{PiDelta4}
\end{eqnarray}
It is convenient to split $P_{i}$ into the part corresponding to
noninteracting particles and the part depending on the interaction, the
result for the former being already known from Eq.\ (\ref{PLZDef}), and to
introduce $\varepsilon $. Thus one obtains at $t\rightarrow \infty $
\begin{equation}
P_{i}\cong 1-\varepsilon +\varepsilon ^{2}/2+\varepsilon
^{2}\sum_{j}I_{ij}\cong e^{-\varepsilon }+\varepsilon ^{2}\sum_{j}I_{ij}
\label{PiExpansion}
\end{equation}
with
\begin{eqnarray}
I_{ij} &=&\left( \frac{v}{2\pi \hbar }\right) ^{2}\left[ \left| \tilde{c}%
_{ij}^{(2)(\mathrm{ni})}(\infty )\right| ^{2}-\left| \tilde{c}%
_{ij}^{(2)}(\infty )\right| ^{2}\right]  \nonumber \\
&&+\left( \frac{v}{2\pi \hbar }\right) ^{2}2\func{Re}\left\{ \tilde{c}%
_{i}^{(1)\ast }(\infty )\int_{-\infty }^{\infty }dte^{i\Phi _{i}(t)}\right.
\nonumber \\
&&\times \left. \left[ \tilde{c}_{ij}^{(2)(\mathrm{ni})}(t)e^{-i\Phi _{ij}^{(%
\mathrm{ni})}(t)}-\tilde{c}_{ij}^{(2)}(t)e^{-i\Phi _{ij}(t)}\right] \right\}
.  \label{IijRes}
\end{eqnarray}
Here the quantities with the superscript $(\mathrm{ni})$ correspond to the
noninteracting system, $J_{ij}=0.$

There is a seeming paradox in the derivation of Eqs.\ (\ref{PiExpansion}), (%
\ref{IijRes}). For a macroscopic system, $N\gg 1,$ even at fast sweep, $%
\varepsilon \ll 1,$ a macroscopic part of the particles $\varepsilon N$ is
flipping out of the starting state with all spins down. However the wave
function of Eq.\ (\ref{PsiForm}) with only one or two spins flipped was used
in the calculation above, whereas the coefficients $c_{ijk}$ etc. have been
dropped. Our calculations is nevertheless correct, because extending of the
expansion of Eq.\ (\ref{cExpansion}) to include more terms would only lead
to contributions of orders higher than $\Delta ^{4}$ or, correspondingly,
higher than $\varepsilon ^{2}$ that we are not interested in. The seeming
paradox has nothing to do with the interaction and it emerges already for a
system of $N$ identical noninteracting TLSs. In this case one can just solve
a one-particle problem exactly and expand $P$ of Eq.\ (\ref{PLZDef}) up to $%
\varepsilon ^{2}$ or, alternatively, one can solve the one-particle problem
perturbatively using lines 1 and 2 of Eq.\ (\ref{cExpansion}). The result is
$P=1-\varepsilon +\varepsilon ^{2}/2+\ldots .$ On the other hand, one can do
the perturbative expansion for the whole system of $N$ particles up to $%
\varepsilon ^{2}$ or higher that evidently leads to the same result. The key
observation is that at any order in $\varepsilon $ (both with and without
the interaction) the terms that are diverging in the limit $N\rightarrow
\infty $ cancel each other, as it should be. In particular, one can see from
Eqs.\ (\ref{PiExpansion}), (\ref{IijRes}) that there is no proplem in the
thermodynamic limit since $I_{ij}$ decays with increasing the distance
between $i$ and $j$ if the interaction $J_{ij}$ does so. In connection to
the above it should be mentioned that the truncated Schr\"{o}dinger
equation, Eq.\ (\ref{SchrEqSlow}) cannot be directly solved numerically. The
problem is that the solution of this equation contains all powers of $\Delta
$ or $\varepsilon ,$ and starting from $\Delta ^{6}$ or $\varepsilon ^{3}$
not all terms of the same order are taken into account. This leads to
noncancellation of terms diverging in the limit $N\rightarrow \infty $
(i.e., spurious terms of order $\varepsilon ^{3}N$ etc.) and thus to an
unphysical behavior of the whole solution.

Calculation of the double and triple time integrals in Eq.\ (\ref{IijRes})
requires essential efforts. At least for the average staying propability of
Eq.\ (\ref{PAvrDef}) where only the symmetrized quantity
\begin{equation}
I_{ij}\Longrightarrow I_{ij}^{(\mathrm{sym})}=\frac{1}{2}\left(
I_{ij}+I_{ji}\right)  \label{IijSymDef}
\end{equation}
is needed and from Eq.\ (\ref{PAvrDef}) one obtains Eq.\ (\ref{PExpansion})
with
\begin{equation}
I_{0}=\frac{1}{N}\sum_{ij}I_{ij}=\frac{1}{N}\sum_{ij}I_{ij}^{(\mathrm{sym})}.
\label{I0Def}
\end{equation}
$I_{ij}^{(\mathrm{sym})}$ can be expressed as
\begin{equation}
I_{ij}^{(\mathrm{sym})}=A_{ij}+\cos \left( 2\pi \tilde{\delta}_{ij}\beta
_{ij}\right) B_{ij}+\sin \left( 2\pi \tilde{\delta}_{ij}\beta _{ij}\right)
C_{ij}.  \label{IDef}
\end{equation}
Here $A_{ij},$ $B_{ij},$ $C_{ij}$ \ are combinations of Fresnel integrals $%
C(x)$ and $S(x)$
\begin{eqnarray}
A_{ij} &=&\frac{1}{2}-\frac{1}{4}\left[ \frac{1}{2}-C\left( \gamma
_{ij}\right) \right] ^{2}-\frac{1}{4}\left[ \frac{1}{2}-S\left( \gamma
_{ij}\right) \right] ^{2}  \nonumber \\
&&-\frac{1}{4}\left[ \frac{1}{2}-C\left( \gamma _{ji}\right) \right] ^{2}-%
\frac{1}{4}\left[ \frac{1}{2}-S\left( \gamma _{ji}\right) \right] ^{2}.
\label{AijRes0}
\end{eqnarray}
\begin{eqnarray}
B_{ij} &=&-\frac{1}{2}\left[ \frac{1}{2}-C\left( \gamma _{ij}\right) \right] %
\left[ \frac{1}{2}-C\left( \gamma _{ji}\right) \right]  \nonumber \\
&&-\frac{1}{2}\left[ \frac{1}{2}-S\left( \gamma _{ij}\right) \right] \left[
\frac{1}{2}-S\left( \gamma _{ji}\right) \right] .  \label{BijRes0}
\end{eqnarray}
\begin{eqnarray}
C_{ij} &=&\frac{1}{2}\left[ \frac{1}{2}-C\left( \gamma _{ij}\right) \right] %
\left[ \frac{1}{2}-S\left( \gamma _{ji}\right) \right]  \nonumber \\
&&-\frac{1}{2}\left[ \frac{1}{2}-S\left( \gamma _{ij}\right) \right] \left[
\frac{1}{2}-C\left( \gamma _{ji}\right) \right] ,  \label{CijRes0}
\end{eqnarray}
and the dimensionless variable $\gamma _{ij}$ is defined by
\begin{eqnarray}
\gamma _{ij} &\equiv &\tilde{\delta}_{ij}+\beta _{ij},\qquad \tilde{\delta}%
_{ij}\equiv \tilde{\alpha}_{i}-\tilde{\alpha}_{j}  \nonumber \\
\tilde{\alpha}_{i} &\equiv &\frac{\tilde{V}_{i}}{\sqrt{2\pi \hbar v}},\qquad
\beta _{ij}\equiv \frac{4J_{ij}}{\sqrt{2\pi \hbar v}}=\frac{4J_{ij}}{\pi
\Delta }\sqrt{\varepsilon }.  \label{gammaijDef1}
\end{eqnarray}
Here $\tilde{V}_{i}$ contains interaction according to its definition in
Eq.\ (\ref{VIntDef}). Note that independent arguments in Eqs.\ (\ref{IDef}%
)--(\ref{CijRes0}) are reduced interaction $\beta _{ij}$ and resonance
shifts $\delta _{ij},$ thus we also will be using explicit notations of the
kind
\begin{equation}
\Phi _{ij}\equiv \Phi (\tilde{\delta}_{ij},\beta _{ij})
\label{ExplicitNotation}
\end{equation}
for any of the functions $I_{ij}^{(\mathrm{sym})},$ $A_{ij},$ $B_{ij},$ and $%
C_{ij}.$ Similarly to Eq.\ (\ref{VIntDef}), one can write
\begin{equation}
\tilde{\delta}_{ij}=\delta _{ij}+\delta _{ij}^{(\mathrm{int})}.
\label{deltaTildeDef}
\end{equation}

Eqs.\ (\ref{PExpansion}) and (\ref{I0Def})--(\ref{gammaijDef1}) is our main
result that is valid for arbitrary interactions $J_{ij}$ and resonance
shifts $V_{i}$. Note that it has a pair structure and thus it can be
verified against the direct numerical solution for the model of two coupled
particles. Analytical form makes its application practically possible:
Triple time integrals of Eq.\ (\ref{IijRes}) cannot be computed numerically
with a reasonable precision within a reasonable time.

\begin{figure}[t]
\unitlength1cm
\begin{picture}(11,6)
\centerline{\psfig{file=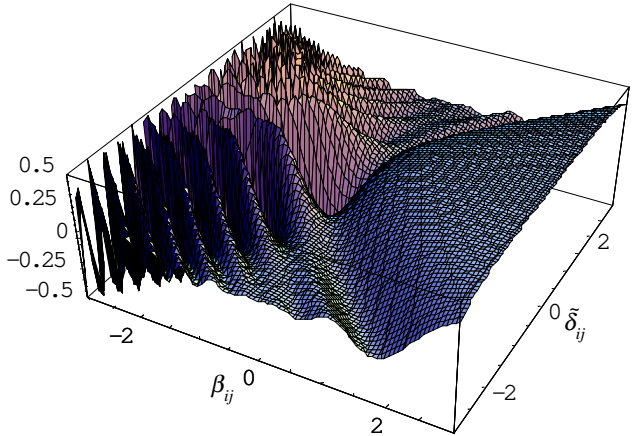,angle=-90,width=9cm}}
\end{picture}
\caption{$I_{ij}^{(\mathrm{sym})}$ vs its independent arguments $\protect%
\beta _{ij}$ and $\tilde{\protect\delta}_{ij}.$}
\label{IijPlot3D}
\end{figure}
\begin{figure}[t]
\unitlength1cm
\begin{picture}(11,6)
\centerline{\psfig{file=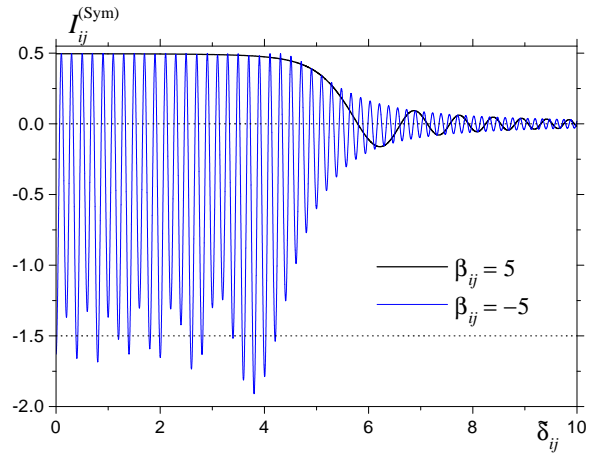,angle=-90,width=9cm}}
\end{picture}
\caption{$I_{ij}^{(\mathrm{sym})}$ as a function of the resonance shift $%
\tilde{\protect\delta}_{ij}$ for $\protect\beta _{ij}=\pm 5.$}
\label{Fig-Iijvsdelta}
\end{figure}

\section{Analysis of the solution}

\label{Sec-Analysis}

\subsection{General properties and limiting cases}

\label{Sec-AnalysisLimiting}

A three-dimensional plot of $I_{ij}^{(\mathrm{sym})}$ vs its independent
arguments $\beta _{ij}$ and $\tilde{\delta}_{ij}$ is given in Fig.\ \ref
{IijPlot3D}. One can see that $I_{ij}^{(\mathrm{sym})}$ has a plateau $%
I_{ij}^{(\mathrm{sym})}\cong 1/2$ for strong ferromagnetic interactions, $%
\beta _{ij}\gg 1.$ This means, according to Eqs.\ (\ref{PExpansion}) and (%
\ref{I0Def})\ ferromagnetic interactions suppress LZ transitions. For strong
antiferromagnetic interactions, $-\beta _{ij}\gg 1,$ the value of $I_{ij}^{(%
\mathrm{sym})}$ oscillates between the upper bound 1/2 and the lower bound
around $-3/2$(see Fig.\ \ref{Fig-Iijvsdelta}), i.e., on average
antiferromagnetic interactions enhance transitions. For large resonance
shifts, $\left| \tilde{\delta}_{ij}\right| \gtrsim \left| \beta _{ij}\right|
,$ the value of $I_{ij}^{(\mathrm{sym})}$ decays to zero. This ``causality''
is physically expected since two particles having resonances at very
different values of the sweep field $W(t)$ do not affect LZ transitions of
each other.

It can be shown that $I_{ij}^{(\mathrm{sym})}$ satisfies the sum rule
\begin{equation}
\int_{-\infty }^{\infty }d\delta I^{(\mathrm{sym})}(\delta ,\beta )=\beta
\label{FundamentalRelation}
\end{equation}
that drastically simplifies the analytical results in the case of strong
static disorder.

In the homogeneous case $\tilde{\delta}_{ij}=0$ one has $I_{ij}^{(\mathrm{sym%
})}=I_{ij}=I(0,\beta _{ij})\equiv F\left( \beta _{ij}\right) ,$ where
\begin{equation}
F\left( \beta \right) =C\left( \beta \right) \left[ 1-C\left( \beta \right) %
\right] +S\left( \beta \right) \left[ 1-S\left( \beta \right) \right] .
\label{IijHomo}
\end{equation}
The limiting forms of $F(\beta )$ are
\begin{equation}
F(\beta )\cong \left\{
\begin{array}{ll}
-\frac{3}{2}-\frac{2\sqrt{2}}{\pi \beta }\cos \left( \frac{\pi }{2}\beta
^{2}+\frac{\pi }{4}\right) , & -\beta \gg 1 \\
\beta -\beta ^{2}, & |\beta |\ll 1 \\
\frac{1}{2}-\frac{1}{\left( \pi \beta \right) ^{2}}, & \beta \gg 1.
\end{array}
\right.  \label{FLims}
\end{equation}
For the weak interaction, $|\beta _{ij}|\ll 1,$ Eq.\ (\ref{PExpansion}) then
yields at the leading order
\begin{equation}
P\cong 1-\varepsilon +\frac{\varepsilon ^{2}}{2}+\frac{4J_{0}}{\pi \Delta }%
\varepsilon ^{5/2},  \label{PJweak}
\end{equation}
a generalization of Eq.\ (26) of Ref.\ \onlinecite{gar03prb} for the
arbitrary form of $J_{ij}.$ Note that Eq.\ (\ref{PJweak}) is essentially a
MFA result (see a separate consideration of the MFA for our model in the
Appendix) as it only depends on the zero Fourier component $J_{0}$ of the
coupling $J_{ij}.$ In contrast to thermodynamic systems, here the
applicability of the MFA is controlled by the strength of the interaction in
addition to its radius. The correction to the mean-field result of Eq.\ (\ref
{PJweak}) is described by the term $-\beta ^{2}$ in the central line of Eq.\
(\ref{FLims}). For the nearest-neighbor interaction with $z$ nearest
neighbors the relative correction to the last term of Eq.\ (\ref{FLims}) is $%
-\left[ 4J_{0}/(\pi \Delta )\right] \varepsilon ^{1/2}/z$ that becomes small
both for large $z$ and for fast sweep.

The saturation for strong ferro- and antiferromagnetic interactions in Eq.\ (%
\ref{FLims}) corresponds to the case of well-separated resonances studied in
Sec. III of Ref.\ \onlinecite{gar03prb}. At fast sweep $\varepsilon \ll 1,$
LZ transitions happen in the range $W\sim W_{LZ}$ around the level crossing,
where\cite{garsch02prb}
\begin{equation}
W_{LZ}=\sqrt{2\pi \hbar v}=\pi \Delta /\sqrt{\varepsilon }.  \label{WLZDef}
\end{equation}
That is, in Eq.\ (\ref{gammaijDef1}) $\beta _{ij}=4J_{ij}/W_{LZ}.$ The
resonances are well separated for $\left| J_{ij}\right| \gg W_{LZ},$ i.e.,
for $\left| \beta _{ij}\right| \gg 1.$ This limit cannot be described by the
MFA. The latter becomes valid in the limit of nonseparated resonances, $%
\left| \beta _{ij}\right| \ll 1.$

Let us proceed to the inhomogeneous case, $\tilde{\delta}_{ij}\neq 0$. For $%
|\beta _{ij}|\ll 1$ Eq.\ (\ref{IDef}) yields
\begin{equation}
I_{ij}^{(\mathrm{sym})}\cong \Xi (\tilde{\delta}_{ij})\beta _{ij}
\label{Iijweak}
\end{equation}
with
\begin{eqnarray}
\Xi (\delta ) &=&\sin \left( \frac{\pi }{2}\delta ^{2}\right) +\cos \left(
\frac{\pi }{2}\delta ^{2}\right) +\pi \delta \left[ S\left( \delta \right)
-C\left( \delta \right) \right]  \nonumber \\
&\cong &\left\{
\begin{array}{cc}
1-\pi \delta ^{2}/2, & \delta ^{2}\ll 1 \\
\frac{\sqrt{2}}{\pi \delta ^{2}}\cos \left( \frac{\pi }{2}\delta ^{2}+\frac{%
\pi }{4}\right) , & \delta ^{2}\gg 1.
\end{array}
\right.  \label{XiDef}
\end{eqnarray}
$\Xi (\delta )$ satisfies the sum rule
\begin{equation}
\int_{-\infty }^{\infty }d\delta \,\Xi (\delta )=1  \label{XiSumRule}
\end{equation}
that is a particular case of the more general Eq.\ (\ref{FundamentalRelation}%
). We will see in the Appendix that Eq.\ (\ref{Iijweak}) also follows from
the MFA.

For $\left| \tilde{\delta}_{ij}\right| -\max (|\beta _{ij}|,1)\gg 1,$ Eq.\ (%
\ref{IDef}) yields a small value
\begin{equation}
I_{ij}^{(\mathrm{sym})}\cong \frac{\sqrt{2}}{\pi }\cos \left( \frac{\pi }{2}%
\gamma _{ji}^{2}+\frac{\pi }{4}\right) \frac{\beta _{ij}}{\tilde{\delta}%
_{ij}^{2}-\beta _{ij}^{2}},  \label{IijStronggamma0}
\end{equation}
as explained above. Note that here $\gamma _{ji}=-\tilde{\delta}_{ij}+\beta
_{ij}.$ For $\left| \tilde{\delta}_{ij}\right| \gg \max (|\beta _{ij}|,1)$
this reduces to Eq.\ (\ref{Iijweak}) with the second limiting form of Eq.\ (%
\ref{XiDef}). That is, the applicability range of Eq.\ (\ref{Iijweak}) is
larger than just $|\beta _{ij}|\ll 1.$

In the case $|\beta _{ij}|-\left| \tilde{\delta}_{ij}\right| \gg 1$ one
obtains
\begin{equation}
I_{ij}^{(\mathrm{sym})}\cong \left\{
\begin{array}{ll}
-1/2-\cos \left( 2\pi \tilde{\delta}_{ij}\beta _{ij}\right) , & \beta _{ij}<0
\\
1/2, & \beta _{ij}>0.
\end{array}
\right.  \label{IijLargebeta}
\end{equation}
For long-range interactions such as the DDI each TLS can strongly interact
with many other TLSs with different strenghts $|\beta _{ij}|\gg 1,$ and the
resonance shifts between different particles can be strong and different, $|%
\tilde{\delta}_{ij}|\gg 1.$ In this case the cosine term in Eq.\ (\ref
{IijLargebeta}) averages out in Eq.\ (\ref{I0Def}).

Neglecting the small value given by Eq.\ (\ref{IijStronggamma0}) and
replacing the cosine term by zero in Eq.\ (\ref{IijLargebeta}) one can
combine the expression for $I_{ij}^{(\mathrm{sym})}$ in the case of both
large arguments $\beta _{ij}$ and $\tilde{\delta}_{ij}$:
\begin{equation}
I_{ij}^{(\mathrm{sym})}\cong \frac{1}{2}\mathrm{sign}\left( \beta
_{ij}\right) \theta \left( |\beta _{ij}|-|\tilde{\delta}_{ij}|\right) .
\label{IijSymLargeArgs}
\end{equation}
Here $\theta (x)$ is the step function. Note that this form satisfies the
sum rule, Eq.\ (\ref{FundamentalRelation}).

The terms with $\cos \left( 2\pi \tilde{\delta}_{ij}\beta _{ij}\right) $ and
$\sin \left( 2\pi \tilde{\delta}_{ij}\beta _{ij}\right) $ in Eq.\ (\ref{IDef}%
) represent the effect of the quantum-mechanical phase in the many-body LZ
effect. The occurence of these terms is due to the possibility to come to a
given final state along different ways. The latter can be seen in Fig.\ \ref
{LZEffect-ManyBody} but is absent for the usual LZ effect, Fig.\ \ref
{LZEffect}. The quantum-mechanical amplitudes corresponding to the different
ways add up with their phases in the expression for the staying probability $%
P$ that causes its oscillations. The minimal model that exhibits this effect
is a dimer of two antiferromagnetically coupled two-level systems with
shifted resonances. \cite{gar04prb} Note that the effect of the
quantum-mechanical phase is totally absent in the mean-field approximation.

\subsection{Effect of static disorder of resonance positions}

\label{Sec-AnalysisDisorder}

As can be seen from Eq.\ (\ref{VIntDef}), the shifts of the resonance
positions $\tilde{V}_{i}$ that enter the final formula via Eq.\ (\ref
{gammaijDef1}) are the sums of the two terms: The original shifts $V_{i}$ of
Eq.\ (\ref{Ham}) and the contribution of the interaction $J_{ij}.$ The
former can arise due to different types of static disorder, including that
induced by nuclear spins\cite{weretal00prl} (see below). The latter are
constant and thus irrelevant ($V_{i}^{(\mathrm{int})}-V_{j}^{(\mathrm{int}%
)}\cong 0$) within the body of the sample for short-range interactions, and
in this case they only affect the particles on the surface. For long-range
interactions such as the DDI, $V_{i}^{(\mathrm{int})}-V_{j}^{(\mathrm{int})}$
in general smoothly varies across the sample, depending on the sample shape.

The effect of static disorder can be accounted for by averaging Eq.\ (\ref
{IDef}) over stochastic values of $\alpha _{i}=\left[ V_{i}/\left( \pi
\Delta \right) \right] \sqrt{\varepsilon }$ with a normalized Gaussian
distribution $\rho _{\alpha }(\alpha )=\left( 2\pi \sigma ^{2}\right)
^{-1/2}\exp \left[ -\alpha ^{2}/\left( 2\sigma ^{2}\right) \right] $ and
quadratic average $\left\langle \alpha _{i}^{2}\right\rangle =\sigma ^{2}.$
The distribution of $\delta _{ij}=\alpha _{i}-\alpha _{j}$ is then given by
the same function with $\sigma ^{2}\Rightarrow 2\sigma ^{2}.$ Averaging over
the static disorder
\begin{eqnarray}
\bar{I}^{(\mathrm{sym})}(\sigma ,\delta ^{(\mathrm{int})},\beta )
&=&\int_{-\infty }^{\infty }\frac{d\delta }{\sqrt{4\pi \sigma ^{2}}}\exp %
\left[ -\frac{\delta ^{2}}{4\sigma ^{2}}\right]  \nonumber \\
&&\qquad \times I^{(\mathrm{sym})}(\delta +\delta ^{(\mathrm{int})},\beta )
\label{AveragingDef}
\end{eqnarray}
can only be done numerically in the general case. There are, however,
particular cases in which it can be done analytically.

In the case of \emph{strong} disorder $\sigma \gg \max (1,\left| \beta
\right| )$, the integral in Eq.\ (\ref{AveragingDef}) converges at $\delta
\ll \sigma ,$ so that one can neglect the exponential, shift the integration
variable and use Eq.\ (\ref{FundamentalRelation}) that yields
\begin{equation}
\bar{I}_{ij}^{(\mathrm{sym})}\cong \frac{1}{2\sqrt{\pi }}\frac{\beta _{ij}}{%
\sigma },\qquad \sigma \gg \max (1,\left| \beta _{ij}\right| ).
\label{ISymAvrsigmaLarge}
\end{equation}
Note that in this limit $\bar{I}_{ij}^{(\mathrm{sym})}$ is independent of $%
\delta _{ij}^{(\mathrm{int})}.$ For $\delta _{ij}^{(\mathrm{int})}=0$ and $%
\sigma \gg 1$ one can use the simplified form of $I_{ij}^{(\mathrm{sym})}$
given by Eq.\ (\ref{IijSymLargeArgs}) to obtain
\begin{equation}
\bar{I}_{ij}^{(\mathrm{sym})}\cong \frac{1}{2}\func{erf}\left( \frac{\beta
_{ij}}{2\sigma }\right) ,\qquad \sigma \gg 1.  \label{ISymAvrsigmaLargeErf}
\end{equation}
Indeed, for $\sigma \sim \left| \beta _{ij}\right| \gg 1$ Eq.\ (\ref
{IijSymLargeArgs}) becomes correct while reducing $\left| \beta _{ij}\right|
$ leads to Eq.\ (\ref{ISymAvrsigmaLarge}).

In the case $\left| \beta \right| \ll 1$ one can use Eq.\ (\ref{Iijweak}),
where the averaged value of $\Xi $ can be found analytically for arbitrary $%
\sigma $ in the case $\delta ^{(\mathrm{int})}=0$. One obtains
\begin{equation}
\bar{I}_{ij}^{(\mathrm{sym})}\cong \beta _{ij}\overline{\Xi }(\sigma
),\qquad \delta ^{(\mathrm{int})}=0,  \label{ISymAvrbetaSmall}
\end{equation}
where $\overline{\Xi }(\sigma )$ is given by
\begin{eqnarray}
\overline{\Xi }(\sigma ) &=&f\left( 2\pi \sigma ^{2}\right) ,\qquad
f(x)\equiv \frac{x}{\sqrt{2(1+x^{2})}}  \nonumber \\
&&\times \left[ \frac{x+1}{\sqrt{\sqrt{1+x^{2}}+1}}-\frac{x-1}{\sqrt{\sqrt{%
1+x^{2}}-1}}\right] .  \label{FAvrSmallbeta}
\end{eqnarray}
Here $f(x)$ monotonically decreases and has limiting forms
\begin{equation}
f(x)\cong \left\{
\begin{array}{cc}
1-x/2, & x\ll 1 \\
1/\sqrt{2x}, & x\gg 1.
\end{array}
\right.  \label{flims}
\end{equation}
The latter limiting form,
\begin{equation}
\overline{\Xi }(\sigma )\cong \frac{1}{2\sqrt{\pi }\sigma },
\label{XiAvrLargesigma}
\end{equation}
follows from Eq.\ (\ref{XiSumRule}) and it is valid for $\delta ^{(\mathrm{%
int})}\neq 0$ as well.

Let us now consider the case of \emph{weak} disorder $\sigma \ll 1,$ setting
$\delta ^{(\mathrm{int})}=0.$ In this case the main effect arises for large
negative $\beta $ [see Eq.\ (\ref{IijLargebeta})], where even small disorder
can average out fast oscillations of $I_{ij}^{(\mathrm{sym})}.$ Combining
Eq.\ (\ref{IijHomo}) with Eq.\ (\ref{IijLargebeta}) and using $\left\langle
\cos \left( 2\pi \delta \beta \right) \right\rangle =e^{-\left( 2\pi \beta
\sigma \right) ^{2}}$ one obtains
\begin{equation}
\bar{I}_{ij}^{(\mathrm{sym})}\cong F(\beta _{ij})+\theta (-\beta
_{ij})\left( 1-e^{-\left( 2\pi \beta _{ij}\sigma \right) ^{2}}\right) .
\label{FAvrDef}
\end{equation}
It can be seen that the limiting value $\lim_{\beta \rightarrow -\infty
}F(\beta )=-3/2$ in Eq.\ (\ref{FLims}) is unstable with respect to small
static disorder$:$ For any $\sigma >0$ one obtains $\lim_{\beta \rightarrow
-\infty }\bar{I}^{(\mathrm{sym})}(\sigma ,0,\beta )=-1/2.$ The reason for
that is the effect of the quantum-mechanical phase that leads to the fast
oscillating terms in Eq.\ (\ref{IDef}).

\begin{figure}[t]
\unitlength1cm
\begin{picture}(11,6)
\centerline{\psfig{file=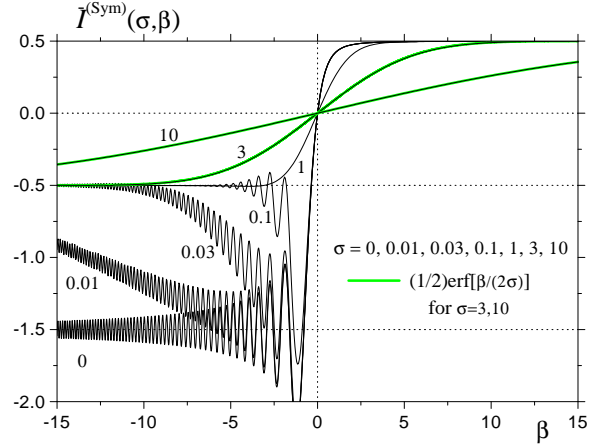,angle=-90,width=9cm}}
\end{picture}
\caption{$\bar{I}_{ij}^{(\mathrm{sym})}$ as a function of $\protect\beta $
for different strengths of static disorder $\protect\sigma .$}
\label{Fig-IAvr}
\end{figure}

$\bar{I}_{ij}^{(\mathrm{sym})}$ averaged over static disorder is shown as a
functiion of $\beta $ for different $\sigma $ in Fig.\ \ref{Fig-IAvr}. One
can see that $\sigma \neq 0$ changes the asymptotic behavior of $\bar{I}%
_{ij}^{(\mathrm{sym})}$ at $\beta \rightarrow -\infty $ and that $\bar{I}%
_{ij}^{(\mathrm{sym})}$ becomes odd in $\beta $ for $\sigma \gg 1$ (in fact,
already for $\sigma \gtrsim 1),$ as described by Eq.\ (\ref
{ISymAvrsigmaLargeErf}).

In molecular magnets distribution of resonance positions for electronic
spins $V_{i}$ is mainly induced by the coupling to nuclear spins. This can
be both the contact hyperfine interaction with the nuclear spins of the
magnetic atoms, as in Fe$_{8}$ containing the isotope $^{57}$Fe and in Mn$%
_{12}$, and the dipole-dipole interaction with magnetic moments of nuclear
spins of non-magnetic atoms, mainly protons. The latter is small as the
nuclear magneton but the number of nonmagnetic atoms interacting with the
magnetic atoms is of order 10$^{2},$ so that the total effect is quite
substantial. Measurements of Ref.\ \onlinecite{weretal00prl} for the
resonance between the ground states $\left| \pm S\right\rangle $ ($S=10$) in
Fe$_{8}$ with the standard Fe isotope (no nuclear spins on Fe atoms) yield a
Gaussian line shape with the width $\sigma _{H}=0.8$ mT that is in agreement
with the theoretical evaluation $\sigma _{H}=0.5$ mT. \cite{weretal00prl}
For the $^{57}$Fe isotope the line width is only about two times larger, 1.2
mT and 1.1 mT, respectively. \cite{weretal00prl} The dimensionless
dispersion $\sigma $ introduced earlier in this section is defined similarly
to Eq.\ (\ref{gammaijDef1}):
\begin{equation}
\sigma =\frac{\sigma _{V}}{\sqrt{2\pi \hbar v}}=\frac{\sigma _{V}}{\pi
\Delta }\sqrt{\varepsilon }\equiv \sigma _{0}\sqrt{\varepsilon },
\label{sigmaDef}
\end{equation}
where $\sigma _{V}=2Sg\mu _{B}\sigma _{H}$ is the dispersion in the energy
units. With $\sigma _{H}=0.8$ mT one obtains $\sigma _{V}/k_{B}=21$ mK. With
$\Delta /k_{B}\simeq 10^{-7}$ K for the ground-state resonance one obtains a
huge value $\sigma _{0}\simeq 0.68\times 10^{5}$ that makes $\sigma $ very
large even for a moderately fast sweep such as $\varepsilon \sim 10^{-2}.$
With the help of Eq.\ (\ref{gammaijDef1}) one can rewrite Eq.\ (\ref
{ISymAvrsigmaLarge}) in the natural form as
\begin{equation}
\bar{I}_{ij}^{(\mathrm{sym})}\cong \frac{2}{\sqrt{\pi }}\frac{J_{ij}}{\sigma
_{V}},\qquad \left| J_{ij}\right| \ll \sigma _{V}
\label{ISymAvrsigmaLargeNatural}
\end{equation}
that is independent of $\Delta .$ One can show that this result follows from
the MFA as well, as all our limiting cases where $\bar{I}_{ij}^{(\mathrm{sym}%
)}\varpropto J_{ij}$. If the disorder is so strong that $\left|
J_{ij}\right| \ll \sigma _{V}$ is satisfied for \emph{all} distances,
including the nearest neighbors, then the many-body LZ effect can be
described by the MFA.

It remains only to justify that nuclear spins can be considered as static
disorder. The appropriate condition is
\begin{equation}
\omega _{LZ}\gg \Gamma _{\mathrm{nuc}},  \label{StaticDisorderCond}
\end{equation}
where $\omega _{LZ}$ the characteristic frequency of the LZ transition and $%
\Gamma _{\mathrm{nuc}}$ is the nuclear relaxation rate. At fast sweep, $%
\varepsilon \ll 1,$ LZ transition occurs in the range of sweep fields
defined by Eq.\ (\ref{WLZDef}). This yields the Landau-Zener time $%
t_{LZ}=W_{LZ}/v=\sqrt{2\pi \hbar /v}$ and the LZ frequency
\begin{equation}
\omega _{LZ}=\frac{1}{t_{LZ}}=\sqrt{\frac{v}{2\pi \hbar }}=\frac{\Delta }{%
2\hbar }\frac{1}{\sqrt{\varepsilon }}.  \label{omegaLZDef}
\end{equation}
For $\Delta /k_{B}\simeq 10^{-7}$ K one obtains $\omega _{LZ}=0.65\times
10^{4}/\sqrt{\varepsilon }$ s$^{-1}$ that grows with the sweep rate. On the
other hand, recent measurements \cite{moretal04prl} on Mn$_{12}$ yield $%
\Gamma _{\mathrm{nuc}}$ in the range between $10^{-1}$ and 10$^{-2}$ s$^{-1}$
for $T<1$ K. Thus nuclear spins are really slow and they can be considered
as static disorder in a wide range of sweep rates. The situation is unlikely
to be strongly different in Fe$_{8}.$

\section{DDI in crystals of ellipsoidal shape or with strong disorder}

\label{Sec-DDI}

Let us now turn to the DDI between tunneling spins $\pm S$ of magnetic
molecules aligned along the easy axis pointing in some direction $\mathbf{m}%
, $ $\left| \mathbf{m}\right| =1$:
\begin{equation}
J_{ij}=E_{D}\phi _{ij},\qquad \phi _{ij}=v_{0}\frac{3\cos ^{2}\theta _{ij}-1%
}{r_{ij}^{3}},  \label{JijDDI}
\end{equation}
where
\begin{equation}
E_{D}\equiv \left( g\mu _{B}S\right) ^{2}/v_{0}  \label{EDDef}
\end{equation}
is the dipolar energy, $v_{0}$ is the unit-cell volume, $r_{ij}$ is the
distance between the sites $i$ and $j$, and $\cos \theta _{ij}=\left(
\mathbf{r}_{ij}\mathbf{\cdot m}\right) /r_{ij}.$ The two mostly well known
molecular magnets are Mn$_{12}$ and Fe$_{8},$ both having the effective spin
$S=10$. Mn$_{12}$ crystallizes in a tetragonal lattice with parameters $%
a=b=17.319$ \AA ,\ $c=12.388$ \AA\ ($c$ is the easy axis) and $%
v_{0}=abc=3716 $ \AA $^{3}$. Fe$_{8}$ has a triclinic lattice with $a=10.52$
\AA\ ($a$ is the easy axis), $b=14.05$ \AA , $c=15.00$ \AA , $\alpha =89.9%
%TCIMACRO{\UNICODE[m]{0xb0}}%
%BeginExpansion
{{}^\circ}%
%EndExpansion
,$ $\beta =109.6%
%TCIMACRO{\UNICODE[m]{0xb0}}%
%BeginExpansion
{{}^\circ}%
%EndExpansion
,$ $\gamma =109.3%
%TCIMACRO{\UNICODE[m]{0xb0}}%
%BeginExpansion
{{}^\circ}%
%EndExpansion
$ and $v_{0}=abc\sin \alpha \sin \beta \sin \gamma =1971$ \AA $^{3}$ (see,
e.g., Ref.\ \onlinecite{marchuaha01}). Fe$_{8}$ is more convenient as a
model system for us as (i) the standard Fe isotope does not have a nuclear
spin that is ignored in our theory and (ii) the splitting $\Delta $ in Fe$%
_{8}$ has a well-defined origin and it can be estimated theoretically. One
can write $\beta _{ij}$ of Eq.\ (\ref{gammaijDef1}) in the form
\begin{equation}
\beta _{ij}=\xi \phi _{ij},\qquad \xi \equiv \xi _{0}\varepsilon
^{1/2},\qquad \xi _{0}\equiv \frac{4E_{D}}{\pi \Delta }.  \label{xiDef}
\end{equation}
For Fe$_{8}$ $E_{D}/k_{B}=126.4$ mK and $\Delta /k_{B}\simeq 10^{-7}$ K, so
that $\xi _{0}\simeq 1.6\times 10^{6}.$ Thus for not too fast sweep, $%
\varepsilon \sim 10^{-2},$ one has $\xi \sim 10^{5}.$ This is also an
estimation for the number of spins $N_{\xi }\sim \xi $ within the distance
\begin{equation}
r_{c}\equiv \left( v_{0}\xi \right) ^{1/3}  \label{rcDef}
\end{equation}
($\simeq 600$ \AA\ for Fe$_{8})$ that strongly interact with a given spin, $%
\left| \beta _{ij}\right| \gtrsim 1$.

As can be seen from Eq.\ (\ref{ISymAvrsigmaLarge}), in the case of strong
static disorder, $\sigma \gg 1$ the value of $\bar{I}_{ij}^{(\mathrm{sym})}$
is strongly reduced if $\left| \beta _{ij}\right| \lesssim \sigma .$ The
corresponding characteristic length is
\begin{equation}
r_{\sigma }\equiv \left( v_{0}\xi /\sigma \right) ^{1/3}\ll r_{c}.
\label{rsigmaDef}
\end{equation}
Within this distance the interaction is still strong, $\bar{I}_{ij}^{(%
\mathrm{sym})}\sim 1$. The estimation for the number of spins stronly
interacting with a given spin is
\begin{equation}
N_{\sigma }\sim \frac{r_{\sigma }^{3}}{v_{0}}=\frac{\xi }{\sigma }=\frac{%
4E_{D}}{\sigma _{V}}.  \label{NsigmaDef}
\end{equation}
For Fe$_{8}$ one obtains $N_{\sigma }\simeq 24$ that is large but much
smaller than the values of order $N_{\xi }\sim 10^{5}$ in the absence of
disorder. For the superstrong disorder,
\begin{equation}
\sigma \gtrsim \xi \text{ \ \ \ \ \ or \ \ \ }\sigma _{V}\gtrsim E_{D},
\label{Superstrong}
\end{equation}
there is no range where the interaction is strong, $r_{\sigma }\lesssim
v_{0}^{1/3}$. For $\sigma _{V}\gg E_{D}$, as commented after Eq.\ (\ref
{ISymAvrsigmaLargeNatural}), the mean-field description of the LZ effect
with interactions becomes valid.

\subsection{Samples of ellipsoidal shape}

\label{Sec-DDIEll}

Consider a macroscopically large specimen of ellipsoidal shape. In this case
the magnetostatic field inside the homogeneously magnetized sample (the
system remains in the vicinity of this state in the LZ effect at fast sweep)
is homogeneous in the bulk of the sample, $\delta ^{(\mathrm{int})}=0$. Thus
one only has to make averaging over static disorder using Eq.\ (\ref
{AveragingDef}). According to Eq.\ (\ref{IijLargebeta}) $I_{ij}^{(\mathrm{sym%
})}$ does not diverge for $\left| \beta _{ij}\right| \rightarrow \infty .$
For $\xi \gg 1$ and $\sigma \ll \xi $ (i.e., $N_{\sigma }\gg 1$) one can
replace the sum in Eq.\ (\ref{I0Def}) by an integral converging at $r\equiv
r_{ij}\sim \min (r_{c},r_{\sigma })$ that makes the result independent of
the lattice structure$:$%
\begin{equation}
I_{0}=\int \frac{d^{3}r}{v_{0}}\bar{I}^{(\mathrm{sym})}(\sigma ,\beta (%
\mathbf{r})),  \label{I0Integral}
\end{equation}
where
\begin{equation}
\beta (\mathbf{r})=\xi \phi (\mathbf{r}),\qquad \phi (\mathbf{r})=v_{0}\frac{%
3\cos ^{2}\theta -1}{r^{3}}.  \label{betarDef}
\end{equation}

To simplify the integration in Eq.\ (\ref{I0Integral}), it is convenient to
change the variables and integrate over the direction $\mathbf{r}/r$ and the
value of $\beta $ instead of integrating over the direction $\mathbf{r}/r$
and the distance $r.$ At large distances where $\beta \ll 1,$ one can use
Eq.\ (\ref{ISymAvrbetaSmall}) that makes account of the static disorder.
Since $\bar{I}^{(\mathrm{sym})}(\sigma ,\beta (\mathbf{r}))\varpropto \beta (%
\mathbf{r})$ behaves as the DDI, the result of the integration depends on
the sample shape. Thus one has to be cautious with changing integration
variables. One of possible ways to tackle this problem is to introduce a
subtraction function, say
\begin{equation}
F_{\mathrm{subtr}}(\sigma ,\beta (\mathbf{r}))=\frac{\beta (\mathbf{r})}{%
1+\beta ^{2}(\mathbf{r})}\overline{\Xi }(\sigma )  \label{FSubtr}
\end{equation}
that has a sufficiently simple form, does not diverge at small distances, $%
\beta \rightarrow \infty ,$ and has the same behavior as $I_{ij}^{(\mathrm{\
sym})}$ at large distances, $\beta \ll 1.$ (Other types of subtraction
functions differing by the cutoff at $\beta \rightarrow \infty ,$ for
instance, the function having $1+\left| \beta (\mathbf{r})\right| $ in the
denominator, yield the same results.) With the help of Eq.\ (\ref{FSubtr})
one can write the integral in the form
\begin{eqnarray}
&&I_{0}=\int \frac{d^{3}r}{v_{0}}\left[ \bar{I}^{(\mathrm{sym})}(\sigma
,\beta (\mathbf{r}))-F_{\mathrm{subtr}}(\sigma ,\beta (\mathbf{r}))\right]
\nonumber \\
&&\qquad \qquad +\int \frac{d^{3}r}{v_{0}}F_{\mathrm{subtr}}(\sigma ,\beta (%
\mathbf{r})).  \label{Subtr}
\end{eqnarray}
The first integral converges fast at large distances, thus it does not
depend on the shape and can be rearranged by changing variables as said
above. After doing that, the contribution of $F_{\mathrm{subtr}}$ into this
integral disappears because of the antisymmetry $F_{\mathrm{subtr}}(-\beta
,\sigma )=-F_{\mathrm{subtr}}(\beta ,\sigma ).$ The second integral can be
calculated analytically for the actual sample shape using the results of the
magnetostatics, without changing variables. The final result has the form \
\begin{equation}
I_{0}\cong G\xi ,\qquad G=G_{I}+G_{\mathrm{subtr}}.  \label{GDef}
\end{equation}
Here
\begin{equation}
G_{I}=\frac{8\pi }{9\sqrt{3}}\mathcal{P}\int_{-\infty }^{\infty }d\beta
\frac{\bar{I}^{(\mathrm{sym})}(\sigma ,\beta )}{\beta ^{2}}  \label{GIDef}
\end{equation}
where $\bar{I}^{(\mathrm{sym})}(\sigma ,\beta )$ is given by Eq.\ (\ref
{AveragingDef}). The subtraction contribution reads
\begin{equation}
G_{\mathrm{subtr}}=\left[ K+\left( 1/3-n^{(\mathbf{m})}\right) 4\pi \right]
\overline{\Xi }(\sigma ),  \label{GsubtrDef}
\end{equation}
where the constant $K$ is given by
\begin{equation}
K\equiv -\frac{8\pi }{9}\left( 1-\frac{1}{\sqrt{3}}\ln \frac{\sqrt{3}+1}{%
\sqrt{3}-1}\right) =-0.66924,  \label{KDef}
\end{equation}
and $n^{(\mathbf{m})}$ is the demagnetization coefficient depending on the
sample shape and on the direction of the easy axis (i.e., the magnetization)
$\mathbf{m.}$ If $\mathbf{m}$ coinsides with the $z$ axis, the symmetry axis
of the ellipsoids of revolution, then $n^{(\mathbf{m})}=n^{(z)}=1/3,$ 0, and
1 for a sphere, needle and disc, respectively. In Eq.\ (\ref{GsubtrDef}) the
term with $K$ is the result of the actual calculation for the sphere,
whereas the remainder is known from the magnetostatics.

For $\sigma \ll 1$ we use Eq.\ (\ref{FAvrDef}) to calculate $G_{I}.$ One
obtains
\begin{equation}
G\cong -5.734+16\left( \pi /3\right) ^{5/2}\sigma +\left( 1/3-n^{(\mathbf{m}%
)}\right) 4\pi .  \label{GSmalldeltaalpha}
\end{equation}
Here the linear-$\sigma $ contribution stems from the second term of Eq.\ (%
\ref{FAvrDef}) that changes the asymptotic behavior of $\bar{I}^{(\mathrm{sym%
})}(\sigma ,\beta )$ at $\beta \rightarrow -\infty .$ The large numerical
factor in front of this contribution makes $G$ very sensitive to $\sigma .$

For $\sigma \gg 1$ the integrand of in Eq.\ (\ref{GIDef}) is of order $%
1/\sigma $ and, in addition, it is nearly odd in $\beta .$ Thus the result
of the integration is smaller than $1/\sigma .$ In the limit $\sigma \gg 1$
the leading term in $G$ is $G_{\mathrm{subtr}}$. From Eqs.\ (\ref{GsubtrDef}%
) and (\ref{flims}) one obtains
\begin{equation}
G\cong \frac{K+\left( 1/3-n^{(\mathbf{m})}\right) 4\pi }{2\sqrt{\pi }\sigma }%
.  \label{GLargedeltaalpha}
\end{equation}
Using Eqs.\ (\ref{sigmaDef}), (\ref{GDef}), and (\ref{xiDef}), one obtains
\begin{equation}
I_{0}\cong \frac{2}{\sqrt{\pi }}\left[ K+\left( 1/3-n^{(\mathbf{m})}\right)
4\pi \right] \frac{E_{D}}{\sigma _{V}}  \label{I0StrongDisorder}
\end{equation}
in the case of strong static disorder. For the ground-state resonance in Fe$%
_{8}$ the values of $I_{0}$ are $23.4$, $-4.44,$ $-60.0$ for the needle,
sphere, and disc, respectively. $I_{0}$ changes its sign for the critical
value of the demagnetizing coefficient defined by
\begin{equation}
n_{c}^{(\mathbf{m})}=\frac{K}{4\pi }+\frac{1}{3}=0.280.  \label{nDemagcDef}
\end{equation}
One has $I_{0}>0$ for $n^{(\mathbf{m})}<n_{c}^{(\mathbf{m})}$ and $I_{0}<0$
for $n^{(\mathbf{m})}>n_{c}^{(\mathbf{m})}.$

\begin{figure}[t]
\unitlength1cm
\begin{picture}(11,6)
\centerline{\psfig{file=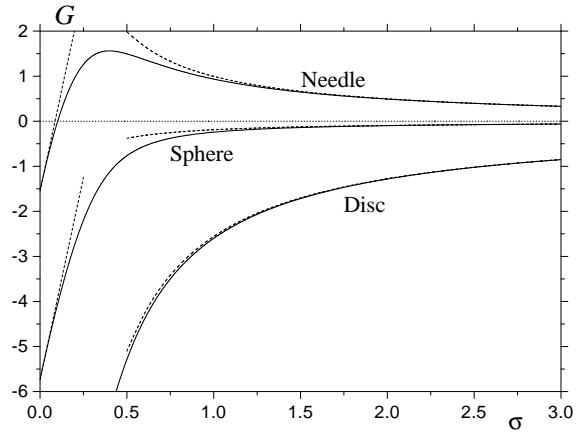,angle=-90,width=9cm}}
\end{picture}
\caption{$G$ of Eq.\ (\ref{GDef}) for the sphere vs width of distribution of
individual resonances $\protect\sigma $. Dashed lines on the left and right
are asymptotes of Eqs.\ (\ref{GSmalldeltaalpha}) and (\ref{GLargedeltaalpha}%
), respectively. }
\label{Fig-LZQ-G}
\end{figure}

The results of the numerical evaluation of $G$ for different values of $%
\sigma $ are shown in Fig.\ \ref{Fig-LZQ-G}. Whereas $G<0$ for the sphere
and disc, DDI acting predominantly antiferromagnetically and enhancing LZ
transitions, the result for the needle in Eq.\ (\ref{GDef}) becomes positive
already for $\sigma \gtrsim 0.1$.

\subsection{Superstrong disorder -- the mean-field limit}

\label{Sec-DDISuperStrong}

In the MFA, $\bar{I}_{ij}^{(\mathrm{sym})}$ is given by Eq.\ (\ref{Iijweak})
that yields Eq.\ (\ref{ISymAvrbetaSmall}) after averaging over static
disorder. We have seen above that the MFA becomes valid for the description
of the LZ effect only in the case of weak interactions $\left| J_{ij}\right|
\ll \Delta /\sqrt{\varepsilon }$or in the case of superstrong disorder $%
\sigma _{V}\gg \max (\left| J_{ij}\right| )$ if the interaction is strong.
In molecular magnets the applicability condition for the MFA is not
fulfilled. Yet we consider it for the sake of completeness. For the DDI one
obtains
\begin{equation}
I_{0}^{(\mathrm{MFA})}=\frac{1}{2\sqrt{\pi }}\frac{\xi }{\sigma }D_{\mathbf{%
mm}}=\frac{2}{\sqrt{\pi }}\frac{E_{D}}{\sigma _{V}}D_{\mathbf{mm}},
\label{I0MFA}
\end{equation}
where $D_{\mathbf{mm}}\equiv $ $\mathbf{m\cdot D}_{\mathbf{m}}$%
\begin{equation}
D_{\mathbf{mm}}\equiv \sum_{j}\phi _{ij}=D_{\mathbf{mm}}^{(\mathrm{Sphere}%
)}+\left( 1/3-n^{(\mathbf{m})}\right) 4\pi   \label{DzzDef}
\end{equation}
and $\mathbf{D}_{\mathbf{m}}$ is the dimensionless quantity proportional to
the dipolar field
\begin{equation}
\mathbf{H}^{(D)}=\frac{g\mu _{B}S}{v_{0}}\mathbf{D}_{\mathbf{m}}
\label{HzMDzzRelation}
\end{equation}
created on the lattice site $i$ by all other spins, $j\neq i,$ pointed in
the $\mathbf{m}$ direction$.$ One can see that $D_{\mathbf{mm}}$ replaces
the universal constant $K$ in Eq.\ (\ref{I0StrongDisorder}).

For ellipsoids of revolution with magnetization $\mathbf{m}$ directed along
the symmetry axis, $D_{\mathbf{mm}}$ is independent of the lattice site $i$
in the main part of the sample, except for the vicinity of the boundaries.
Note that the macroscopic field theory (magnetostatics) is insufficient to
obtain Eq.\ (\ref{DzzDef}). To this end, one can introduce a macroscopic
sphere around the site $i.$ The field from the spins at sites $j$ inside
this sphere can be calculated by a direct summation over the lattice and it
yields the first term in Eq.\ (\ref{DzzDef}). The field from the spins
outside this sphere can be calculated macroscopically and it results into
the second term in Eq.\ (\ref{DzzDef}). For a simple cubic lattice $D_{%
\mathbf{mm}}^{(\mathrm{Sphere})}=0$ by symmetry and the result for $D_{zz}$
becomes purely macroscopic. For tetragonal lattices $D_{\mathbf{mm}}^{(%
\mathrm{Sphere})}>0$ if $a=b>c$ and $D_{\mathbf{mm}}^{(\mathrm{Sphere})}<0$
if $a=b<c.$ Direct numerical calculation yields $D_{\mathbf{mm}}^{(\mathrm{%
Sphere})}=5.139$ for Mn$_{12}$ and $4.072$ for Fe$_{8}.$ Note that $%
E_{0}=-(1/2)D_{zz}E_{D}$ is the dipolar energy per site for the
ferromagnetic spin alignment. Our result $E_{0}=-4.131E_{D}$ for the
needle-shaped Fe$_{8}$ is in qualitative accord with $E_{0}=-4.10E_{D}$ of
Ref.\ \onlinecite{marchuaha01}.

One can see that unlike the rigorous value of $I_{0}$ in the limit $\xi \gg
1,$ the value of $I_{0}^{(\mathrm{MFA})}$ depends on the lattice structure
via $D_{\mathbf{mm}}^{(\mathrm{Sphere})}$. For Mn$_{12}$ Eq.\ (\ref
{nDemagcDef}) with $K\Rightarrow D_{\mathbf{mm}}^{(\mathrm{Sphere})}$ yields
$n_{c}^{(\mathbf{m})}=0.742$ (i.e., $I_{0}^{(\mathrm{MFA})}>0$ for $n^{(%
\mathbf{m})}<0.742),$ whereas for Fe$_{8}$ one obtains $n_{c}^{(\mathbf{m}%
)}=0.657.$ In both cases it essentially differs from the prediction of our
rigorous theory, Eq.\ (\ref{nDemagcDef}).

\subsection{Samples of general shape with strong disorder}

\label{Sec-DDIstrongDisorder}

The strong-disorder result of Eq.\ (\ref{I0StrongDisorder}) is valid for the
samples of nonellipsoidal shape as well. Indeed, the derivation of Eq.\ (\ref
{I0StrongDisorder}) is based on the sum rule, Eq.\ (\ref{FundamentalRelation}%
), that makes the integrand in Eq.\ (\ref{GIDef}) linear in $\beta .$ This
alone would be, however, insufficient, as the limits $\sigma \rightarrow
\infty $ and $\beta \rightarrow \infty $ (i.e., $r\rightarrow 0$) are not
interchangeable. Eq.\ (\ref{ISymAvrsigmaLarge}) diverges at $r\rightarrow 0$
and it needs an appropriate regularization, whereas the true $\bar{I}_{ij}^{(%
\mathrm{sym})}$ does not diverge at $r\rightarrow 0.$ The regularization we
have used above consists in choosing $\beta $ as the integration variable.
Then approximate oddness of $\bar{I}_{ij}^{(\mathrm{sym})}$ in $\beta $ for
large $\sigma $ strongly reduces the integral over $\beta $ and makes it of
order $1/\sigma ^{3}$ that can be neglected. This oddness is also preserved
in the limit $\beta \rightarrow \infty $ for large but fixed $\sigma ,$ as
follows from Eq.\ (\ref{IijSymLargeArgs}). The result of the regularization
at $r\rightarrow 0$ is the nontrivial constant $K$ in Eq.\ (\ref
{I0StrongDisorder}). One can see from Eq.\ (\ref{IijSymLargeArgs}) that $%
\bar{I}_{ij}^{(\mathrm{sym})}$ remains approximately odd in $\beta $ for
large $\beta $ and $\sigma $ even in the case $\delta ^{(\mathrm{int})}\neq
0 $ for nonellipsoidal shapes, especially as $\delta ^{(\mathrm{int}%
)}\rightarrow 0$ for $r\rightarrow 0.$ Thus the situation in the
strong-disorder case $\sigma \gg 1$ is the same for ellipsoidal and
nonellipsoidal sample shapes. In both cases one switches to the integration
over $\beta $ and obtains zero because the integrand is odd. Recalling that
integration over $\beta $ is incompatible with taking into account the
sample shape, one can correct the situation by introducing, e.g., the
subtraction function of Eq.\ (\ref{FSubtr}).

The general result thus can be written in the form
\begin{equation}
I_{0}=\int \frac{d^{3}r}{V}\int \frac{d^{3}r^{\prime }}{v_{0}}\frac{%
\overline{\Xi }(\sigma )\beta (\mathbf{r^{\prime }-r})}{1+\beta ^{2}(\mathbf{%
r^{\prime }-r})},  \label{I0FinalLargesigma}
\end{equation}
that contains the additional averaging over the sample, $\overline{\Xi }%
(\sigma )$ being given by Eq.\ (\ref{XiAvrLargesigma}). For the spherical
shape the result of the integration is known from Eq.\ (\ref
{I0StrongDisorder}), $I_{0}=\xi \overline{\Xi }(\sigma )K.$ The difference
between the general shape and the sphere arises from the integration at
large distances, where $\beta ^{2}$ in the denominator of Eq.\ (\ref
{I0FinalLargesigma}) can be neglected. Thus with the use of Eq.\ (\ref
{betarDef}) one can write
\begin{equation}
I_{0}=\xi \overline{\Xi }(\sigma )\left[ K+\int \frac{d^{3}r}{V}F(\mathbf{r})%
\right] ,  \label{I0withF}
\end{equation}
where the difference $F(\mathbf{r})$ is given by
\begin{eqnarray}
F(\mathbf{r}) &=&\int_{\mathrm{Sample}}\frac{d^{3}r^{\prime }}{v_{0}}\phi (%
\mathbf{r^{\prime }-r})-\int_{\mathrm{Sphere}}\frac{d^{3}r^{\prime }}{v_{0}}%
\phi (\mathbf{r^{\prime }-r})  \nonumber \\
&=&\mathcal{D}_{\mathbf{mm}}(\mathbf{r})-\mathcal{D}_{\mathbf{mm}}^{(\mathrm{%
Sphere})}.  \label{FI0Def}
\end{eqnarray}
Here $\mathcal{D}_{\mathbf{mm}}(\mathbf{r})$ is the $\mathbf{m}$-component
of the reduced macroscopic internal field created by the sample magnetized
in the $\mathbf{m}$ direction,
\begin{equation}
H_{\mathbf{mm}}(\mathbf{r})=\left( g\mu _{B}S/v_{0}\right) \mathcal{D}_{%
\mathbf{mm}}(\mathbf{r}),  \label{DCalzzDef}
\end{equation}
and $\mathcal{D}_{\mathbf{mm}}^{(\mathrm{Sphere})}=(2/3)4\pi .$ Inserting
Eq.\ (\ref{FI0Def}) into Eq.\ (\ref{I0withF}) yields
\begin{equation}
I_{0}\cong \frac{2}{\sqrt{\pi }}\left[ K+\left( 1/3-n^{(\mathbf{m})}\right)
4\pi \right] \frac{E_{D}}{\sigma _{V}}  \label{I0StrongDisorderNEll}
\end{equation}
[c.f. Eq.\ (\ref{I0StrongDisorder})], where the average demagnetization
coefficient is given by
\begin{equation}
\bar{n}^{(\mathbf{m})}=1-\frac{1}{4\pi }\int \frac{d^{3}r}{V}\mathcal{D}_{%
\mathbf{mm}}(\mathbf{r}).  \label{nzbarDef}
\end{equation}
For the needle, sphere, and disc with $\mathbf{m}$ along the symmetry axis
one has $\mathcal{D}_{\mathbf{mm}}=4\pi ,$ $(2/3)4\pi ,$ and 0,
respectively, so that Eq.\ (\ref{nzbarDef}) recovers the known results for
these geometries.

\begin{figure}[t]
\unitlength1cm
\begin{picture}(11,6)
\centerline{\psfig{file=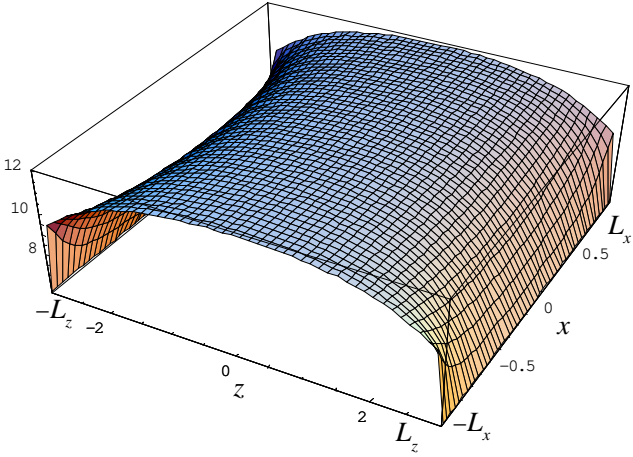,angle=-90,width=9cm}}
\end{picture}
\caption{Dimensionless macroscopic internal field $\mathcal{D}_{\mathbf{mm}}$
of Eq.\ (\ref{DzzArctan}) in the sample of the rectangular shape with $%
L_{x}=L_{y}=1$ and $L_{z}=3,$ uniformly magnetized along the $z$ axis,
plotted vs $x$ and $z$ for $y=0.$}
\label{Fig-LZQ-DipolarField}
\end{figure}

For the general shape the integrals in Eq.\ (\ref{FI0Def}) can be calculated
directly with the use of
\begin{equation}
\phi (\mathbf{r})=\mathbf{m}\cdot v_{0}\frac{3\mathbf{n}_{\mathbf{r}}(%
\mathbf{m}\cdot \mathbf{n}_{\mathbf{r}})-\mathbf{e}_{z}}{r^{3}}=\mathbf{m}%
\cdot v_{0}\mathrm{rot}\frac{\left[ \mathbf{m}\times \mathbf{n}_{\mathbf{r}}%
\right] }{r^{2}}  \label{phivecDef}
\end{equation}
where $\mathbf{n}_{\mathbf{r}}\equiv \mathbf{r/}r.$ The integral formula
\begin{equation}
\int_{V}dV\mathrm{rot}\mathbf{F=}\int_{S}d\mathbf{S}\times \mathbf{F}
\label{rotFormula}
\end{equation}
for an arbitrary vector function $\mathbf{F}(\mathbf{r})$ then reduces the
integration to the surface. Alternatively one can use the the Bio-Savard
formula
\begin{equation}
\mathbf{H}(\mathbf{r})=\frac{1}{c}\int d^{3}r^{\prime }\frac{\mathbf{j}%
\times \left( \mathbf{r-r}^{\prime }\right) }{\left| \mathbf{r-r}^{\prime
}\right| ^{3}}  \label{BioSavard}
\end{equation}
with $\mathbf{j=}c\,\mathrm{rot\,}\mathbf{M}$ and $\mathbf{M=}\left( g\mu
_{B}S/v_{0}\right) \mathbf{m,}$ $\left| \mathbf{m}\right| =1.$ As the
molecular currents $\mathbf{j}$ are nonzero on the surface only, one obtains
\begin{equation}
\mathbf{H}_{\mathbf{m}}(\mathbf{r})=\frac{g\mu _{B}S}{v_{0}}\int_{S}\left[ d%
\mathbf{S}^{\prime }\times \mathbf{m}\right] \times \frac{\mathbf{r^{\prime
}-r}}{\left| \mathbf{r^{\prime }-r}\right| ^{3}}.  \label{BioSavard2}
\end{equation}

For the samples of the box shape with sides $2L_{x},$ $2L_{y}$ and $2L_{z}$
with generally directed $\mathbf{m}$\ the integration in Eq.\ (\ref
{BioSavard2}) yields the sum of 24 arctan terms for the symmetric part of $%
\mathcal{D}_{\mathbf{mm}}(\mathbf{r})$ that makes the contribution into the
volume average in Eq.\ (\ref{nzbarDef})
\begin{eqnarray}
&&\mathcal{D}_{\mathbf{mm}}^{(\mathrm{sym})}(\mathbf{r})  \nonumber \\
&=&(1-m_{x}^{2})\sum_{\eta _{x},\eta _{y},\eta _{z}=\pm 1}\arctan  \nonumber
\\
&&\frac{\left( L_{x}+\eta _{x}x\right) ^{-1}\left( L_{y}+\eta _{y}y\right)
\left( L_{z}+\eta _{z}z\right) }{\sqrt{\left( L_{x}+\eta _{x}x\right)
^{2}+\left( L_{y}+\eta _{y}y\right) ^{2}+\left( L_{z}+\eta _{z}z\right) ^{2}}%
}  \nonumber \\
&&+\left( xyz\Rightarrow yzx\right) +\left( yzx\Rightarrow zxy\right) .
\label{DzzArctan}
\end{eqnarray}
If $\mathbf{m}$ is directed along one of the symmetry axes of the box, the
antisymmetric terms in $\mathcal{D}_{\mathbf{mm}}(\mathbf{r})$ disappear and
this formula yields $\mathcal{D}_{\mathbf{mm}}(\mathbf{r}).$ The result
above is illustrated in Fig.\ \ref{Fig-LZQ-DipolarField} for $m_{z}=1.$
Adopting Eq.\ (\ref{DzzArctan}) in Eq.\ (\ref{nzbarDef}) yields $\bar{n}%
^{(z)}=1/3$ for the cube and $\bar{n}^{(z)}\simeq 0.102$ for the box with
proportions $L_{x}:L_{y}:L_{z}=1:5:8$ that is close to the shape of the
crystal used in Ref.\ \onlinecite{weretal00epl}.

\section{Comparison with experiment}

\label{Sec-experiment}

\begin{figure}[t]
\unitlength1cm
\begin{picture}(11,6)
\centerline{\psfig{file=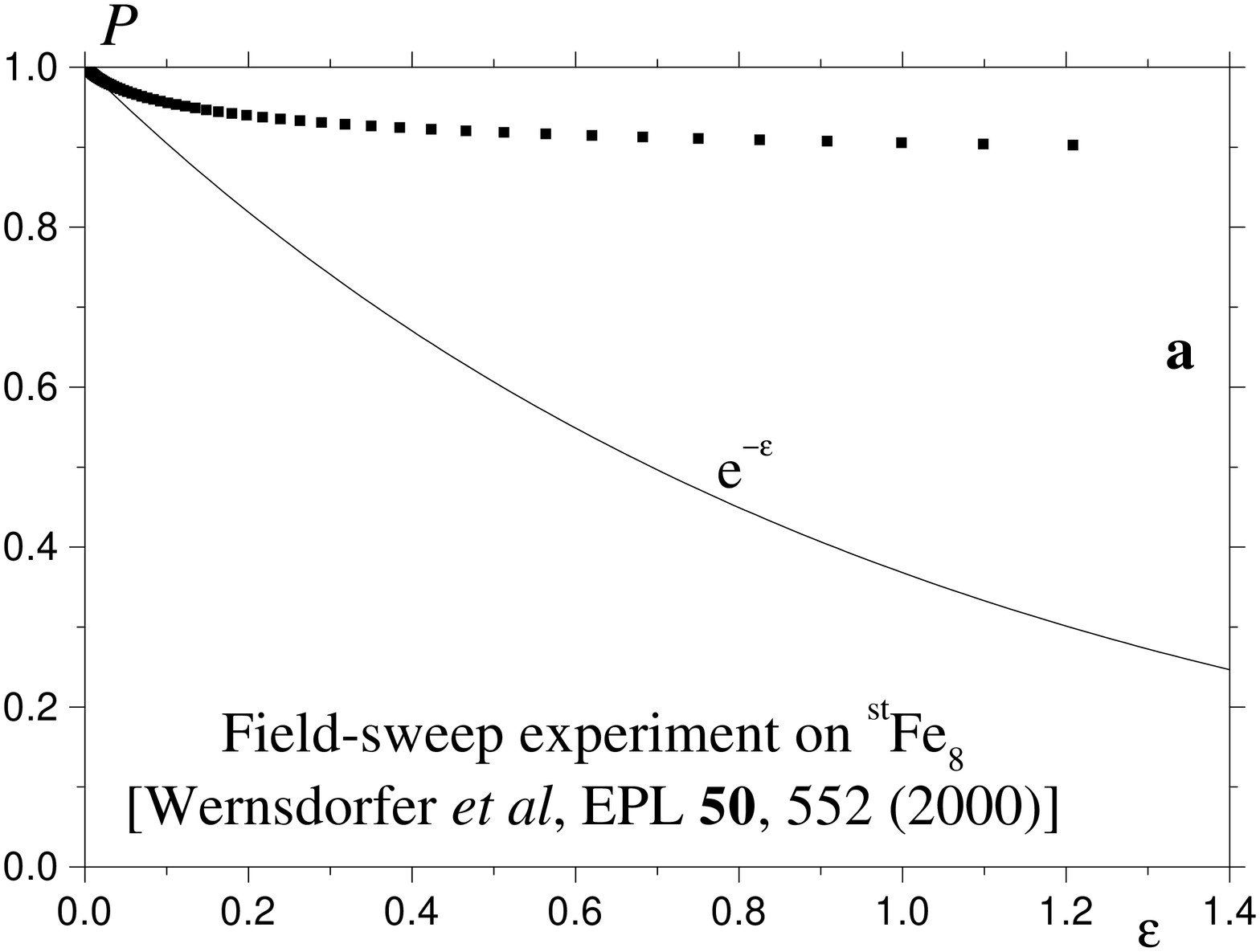,angle=-90,width=9cm}}
\end{picture}
\begin{picture}(11,6)
\centerline{\psfig{file=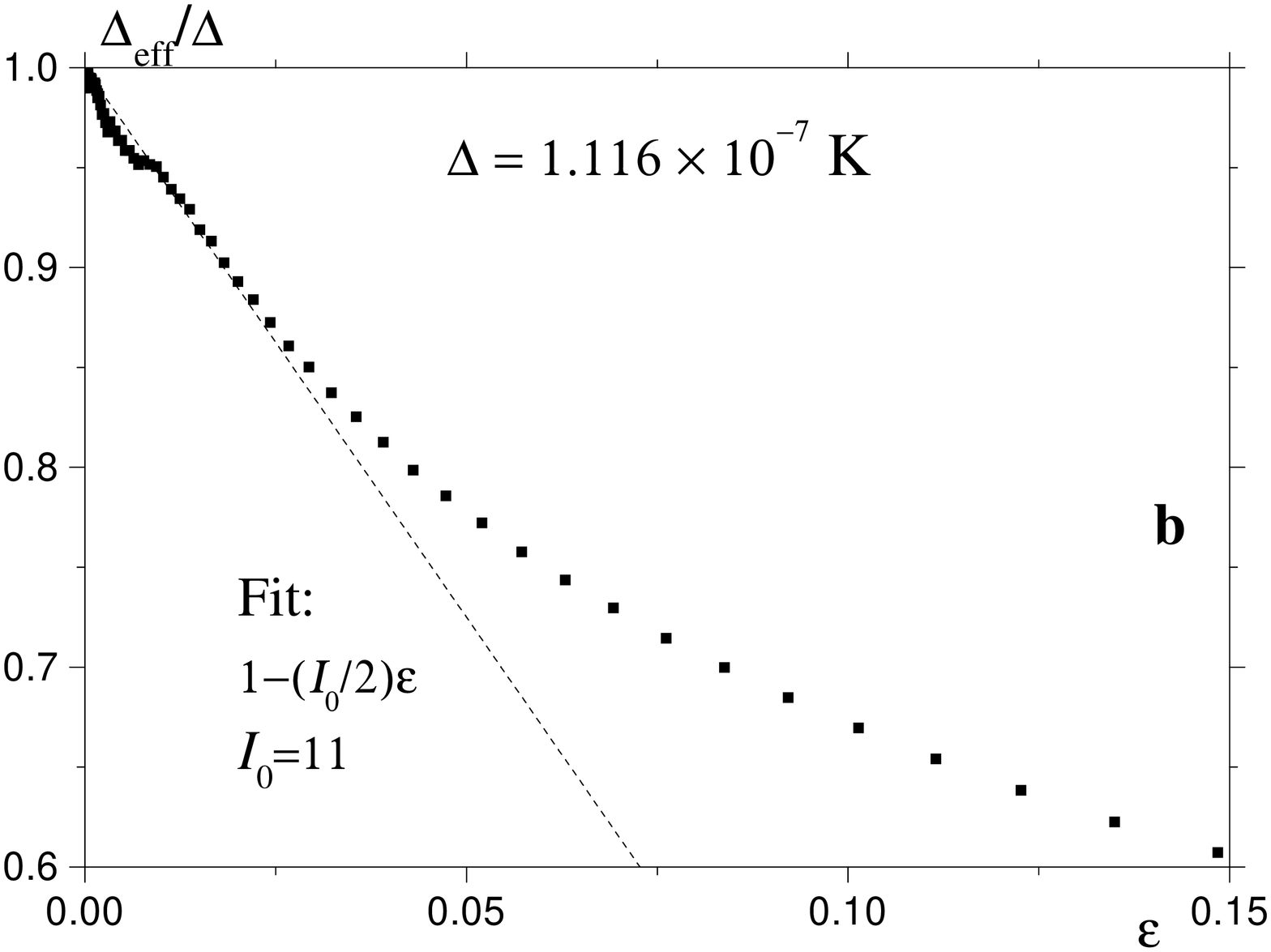,angle=-90,width=9cm}}
\end{picture}
\caption{Replotting experimental results of Ref.\ \onlinecite{weretal00epl}
vs $\protect\varepsilon .$ (a) Staying probability $P.$ \ (b) $\Delta ^{%
\mathrm{eff}}$ defined by Eq.\ (\ref{DeltaeffDef}). Fitting with Eq.\ (\ref
{DeltaeffExpansion}) at $\protect\varepsilon \ll 1$ yields $I_{0}\simeq 11.$}
\label{Fig-Fe8-Wernsdorfer}
\end{figure}

In the sweeping experiments \cite{weretal00epl} studying the $\pm S$
transitions in Fe$_{8}$ the standard LZ effect can be seen down to $%
dB/dt\sim 0.01$ T/s. Using $v=2Sg\mu _{B}dB/dt$ and $\varepsilon =\pi \Delta
^{2}/(2\hbar v)$ with $\Delta /k_{B}\simeq 10^{-7}$ K one obtains the range $%
\varepsilon \lesssim 10^{-2}$ for the standard LZ effect. In the region of
slower sweep $dB/dt\lesssim 0.01$\ T/s (i.e., $\varepsilon \gtrsim 10^{-2}$)
the effective splitting $\Delta ^{\mathrm{eff}}$ calculated from Eq.\ (\ref
{DeltaeffDef})\ goes down. This suggests that here the Landau-Zener effect
is strongly modified by interactions that suppress transitions. It is very
instructive to replot $\exp $erimental data\cite{werpriv} as $P$ vs $%
\varepsilon ,$ see Fig.\ \ref{Fig-Fe8-Wernsdorfer}a. The high plateau of $%
P(\varepsilon )$ at large $\varepsilon $ is similar to that for the spin-bag
model with the ferromagnetic coupling and large $N,$ see, e.g., Fig.\ 3 of
Ref. \onlinecite{gar03prb}. One can estimate $I_{0}$ if one replots the
experimental data for $\Delta ^{\mathrm{eff}}$ vs $\varepsilon $ using Eq.\ (%
\ref{DeltaeffExpansion}), see Fig.\ \ref{Fig-Fe8-Wernsdorfer}. One can see
that $\Delta ^{\mathrm{eff}}$ is apparently linear at small $\varepsilon $.
The fit ignoring the downward bump near the origin yields $I_{0}^{\exp
}\simeq 11.$

The shape of the crystal used in Ref.\ \onlinecite{weretal00epl} was an
elongated platelet ($l_{z}=80$ $\mu $m$,$ $l_{y}=50$ $\mu $m, $l_{x}=10$ $%
\mu $m \cite{werpriv}). The crystallographic easy axis $a$ slightly deviates
from the direction of the longest axis of the crystal, $l_{z}$. It is
rotated by $\alpha =9%
%TCIMACRO{\UNICODE[m]{0xb0}}%
%BeginExpansion
{{}^\circ}%
%EndExpansion
$ in the $yz$ plane and then rotated by $\beta =7%
%TCIMACRO{\UNICODE[m]{0xb0}}%
%BeginExpansion
{{}^\circ}%
%EndExpansion
$ away from the $yz$ plane.\cite{werpriv} Thus for this shape
\begin{eqnarray}
m_{z} &=&\cos \alpha \cos \beta \simeq 0.9803  \nonumber \\
m_{y} &=&\sin \alpha \cos \beta \simeq 0.1553  \nonumber \\
m_{x} &=&\sin \beta \simeq 0.1219.  \label{mWernsdorfer}
\end{eqnarray}
Neglecting this small tilt (i.e., setting $m_{z}=1,$ $m_{x}=m_{y}=0$) and
using Eqs. (\ref{nzbarDef}) and (\ref{DzzArctan}), one obtains $\bar{n}^{(%
\mathbf{m})}=\bar{n}^{(z)}\simeq 0.102.$ Then Eq.\ (\ref
{I0StrongDisorderNEll}) with $E_{D}/k_{B}=126.4$ mK and $\sigma _{V}/k_{B}=21
$ mK yields $I_{0}\simeq 14.8$ that is in a good accord with $I_{0}^{\exp
}\simeq 11.$ Taking into account the tilt of the easy axis improves the
agreement with the experiment: $\bar{n}^{(\mathbf{m})}\simeq 0.113$ and $%
I_{0}\simeq 13.8.$

\section{Discussion}

\label{Sec-discussion}

We have analytically investigated the asymptotic staying probability $P$ in
the many-body Landau-Zener effect at fast sweep, $\varepsilon \ll 1$. The
coefficient $I_{0}$ in the expansion of Eq.\ (\ref{PExpansion}) accounting
for the interaction has been calculated rigorously for arbitrary
interactions $J_{ij}$ and the resonance shifts $V_{i}$ in the
transverse-Ising Hamiltonian of Eq.\ (\ref{Ham}). We have shown that
ferromagnetic interactions ($J>0)$ increase $P$ (i.e., suppress LZ
transitions) whereas antiferromagnetic interactions ($J<0)$ act in the
opposite direction, for the fast sweep.

The resonance shifts $\tilde{V}_{i}$ [see Eq.\ (\ref{VIntDef})] have been
shown to reduce the value of $I_{0}$ since different particles undergoing
transitions at different values of the sweep field $W$ become effectively
decoupled. In all particular cases that we have considered, $I_{0}$ can be
estimated as the \emph{number of neighbors} that are strongly interacting
with a given TLS, $\left| J_{ij}\right| \gtrsim \max (\Delta ,\left| \tilde{V%
}_{i}-\tilde{V}_{j}\right| $) . The sign of $I_{0}$ depends on the sample
shape in the case of the DDI.

Our results differ from those of the mean-field approximation. The MFA
becomes valid for the description of the LZ effect only in the case of weak
interactions $\left| J_{ij}\right| \ll \Delta /\sqrt{\varepsilon }$or in the
case of superstrong disorder $\sigma _{V}\gg \max (\left| J_{ij}\right| )$
if the interaction is strong. Our rigorous expression for $I_{0}$ also
describes models in which MFA predicts irrelevance of the interaction
because of cancellation of molecular fields$.$ Present results can be used
as a ``whetstone'' for checking the quality of different approximations that
can be suggested in the future.

For long-range interactions $J_{ij}$ that exceed $\Delta $ over large
distances, huge values of $I_{0}$ are generated in the absence of the static
disorder. In this case the region of fast sweeps$,$ where a simple
non-interacting LZ effect with $P\cong 1-\varepsilon $ can be observed,
becomes very narrow. The biggest values of $I_{0}$ emerge in samples of
ideal ellipsoidal shape, where the dipolar field is homogeneous$.$

For samples of general shape gradients of the dipolar field reduce the value
of $I_{0}.$ The tunneling particles effectively decouple at the distances
satisfying $\mathbf{r}_{ij}\cdot \nabla \mathcal{D}_{zz}\sim \phi _{ij}.$
Estimating the gradient as $\nabla \mathcal{D}_{zz}\sim L,$ where $L$ is the
linear size of the specimen, one obtains the characteristic length $r_{%
\mathrm{inh}}\equiv \left( Lv_{0}\right) ^{1/4}.$ In the strong-gradient
case $r_{\mathrm{inh}}\lesssim r_{c}$ one can use Eq.\ (\ref{IijSymLargeArgs}%
) that yields $I_{0}\sim r_{\mathrm{inh}}^{3}/v_{0}\sim \left(
V/v_{0}\right) ^{1/4},$ with $V\sim L^{3}.$ This yields still large values
of $I_{0}$ for macroscopic samples. $I_{0}$ even increases if the sample is
elongated or flat because in this case the field becomes more uniform than
in the cube. One obtains $I_{0}\sim -10^{4}$ for the crystal used in LZ
experiments of Ref.\ \onlinecite{weretal00epl}, if one neglects the static
disorder.

Taking into account interactions with nuclear spins considered as frozen-in
disorder drastically reduces $I_{0}$ for Fe$_{8}$ and yields values that
agree with the values extracted from the measurements of Ref.\ %
\onlinecite{weretal00epl} both in sign and magnitude.

It would be important to perform LZ experiments on crystals with a more flat
shape for which the demagnetizing coefficient $n^{(\mathbf{m})}\gtrsim 0.28$
and Eq.\ (\ref{I0StrongDisorder}) yields $I_{0}<0.$ In this case DDI
enhances LZ transitions, and the value of $\Delta _{\mathrm{eff}}$  \emph{%
increases} with $\varepsilon $ starting from its initial value $\Delta ,$
c.f. Fig.\ \ref{Fig-Fe8-Wernsdorfer}. An interesting question is whether
this tendency holds for larger $\varepsilon $ (slower sweep rates) as well.
Our theory is applicable only for small $\varepsilon $ and it cannot answer
this question. It is possible that  $\Delta _{\mathrm{eff}}$ has a maximum
at some $\varepsilon $ and then it falls below $\Delta ,$ as indicated by
our results for some simplified models of the many-body LZ effect.

\section*{Acknowledgment}

The authors thank Wolfgang Wernsdorfer for supplying detailed information on
the Fe$_{8}$ crystal studied in Ref.\ \onlinecite{weretal00epl} and E. M.
Chudnovsky for useful discussions.

\appendix*

\section{Mean-field approximation}

\label{App-MFA}

Let us now consider the mean-field approximation for the many-body LZ effect
at fast sweep and compare its results with our rigorous results obtained
above. Here one has to solve the equations for single spins at different
lattice sites in different effective fields:
\begin{eqnarray}
i\hbar \dot{a}_{i\downarrow } &=&E_{i\downarrow }a_{i\downarrow }-\frac{%
\Delta }{2}a_{i\uparrow }  \nonumber \\
i\hbar \dot{a}_{i\uparrow } &=&E_{i\uparrow }a_{i\uparrow }-\frac{\Delta }{2}%
a_{i\downarrow },  \label{SEMFA}
\end{eqnarray}
where we used the definition of the wave function in the form of Eq.\ (\ref
{PsiPDef}) at each lattice site $i$. Here the energies are
\begin{eqnarray}
E_{i\downarrow } &=&\frac{1}{2}\left[ H_{z}(t)-V_{i}+2\sum_{j}J_{ij}\left%
\langle \sigma _{j}\right\rangle \right]  \nonumber \\
&=&\frac{1}{2}\left[ H_{z}(t)-V_{i}+2\sum_{j}J_{ij}\left( 2\left|
a_{i\uparrow }\right| ^{2}-1\right) \right]  \label{EnergiesMFA}
\end{eqnarray}
and $E_{i\uparrow }=-E_{i\downarrow }.$ Using the transformation similar to
Eq.\ (\ref{PhiDef}) one can rewrite these equations as
\begin{eqnarray}
\frac{d\tilde{a}_{i\downarrow }}{dt} &=&\frac{i\Delta }{2\hbar }e^{i\left[
\Phi _{i\downarrow }(t)-\Phi _{i\uparrow }(t)\right] }\tilde{a}_{i\uparrow }=%
\frac{i\Delta }{2\hbar }e^{i\Phi _{i}(t)}\tilde{a}_{i\uparrow }  \nonumber \\
\frac{d\tilde{a}_{i\uparrow }}{dt} &=&\frac{i\Delta }{2\hbar }e^{-i\Phi
_{i}(t)}\tilde{a}_{i\downarrow }  \label{SEMFASlow}
\end{eqnarray}
with
\begin{equation}
\Phi _{i}(t)=\Phi _{i}^{(0)}(t)+\delta \Phi _{i}(t),  \label{PhiiMFADef}
\end{equation}
where
\begin{equation}
\Phi _{i}^{(0)}(t)=\frac{1}{\hbar }\left[ \frac{vt^{2}}{2}-\tilde{V}_{i}t%
\right] ,  \label{Phi0Def}
\end{equation}
$\tilde{V}_{i}$ is defined by Eq.\ (\ref{VIntDef}), and
\begin{equation}
\delta \Phi _{i}(t)=\frac{4}{\hbar }\sum_{j}J_{ij}\int_{0}^{t}dt\left|
a_{j\uparrow }(t)\right| ^{2}.  \label{deltaPhi0}
\end{equation}
Now we can solve these equations for fast sweep $\varepsilon \ll 1$
iteratively in $\Delta $, similarly to Eq.\ (\ref{cExpansion}) writing
\begin{eqnarray}
\tilde{a}_{i\downarrow } &=&1+\left( \frac{i\Delta }{2\hbar }\right) ^{2}%
\tilde{a}_{i\downarrow }^{(2)}+\ldots  \nonumber \\
\tilde{a}_{i\uparrow } &=&\frac{i\Delta }{2\hbar }\tilde{a}_{i\uparrow
}^{(1)}+\left( \frac{i\Delta }{2\hbar }\right) ^{3}\tilde{a}_{i\uparrow
}^{(3)}+\ldots  \label{aMFAExpansion}
\end{eqnarray}
Further one has to expand $e^{i\Phi _{i}(t)}$ in Eq.\ (\ref{SEMFASlow}) in $%
\left| a_{j\uparrow }(t)\right| ^{2}$ that is small as $\varepsilon $ for
fast sweep:
\begin{equation}
e^{i\Phi _{i}(t)}=e^{i\Phi _{i}^{(0)}(t)}\left[ 1+i\delta \Phi
_{i}(t)+\ldots \right] .  \label{PhiMFAExpExpansion}
\end{equation}
Using Eq.\ (\ref{gammaijDef1}) one can show that \ $\delta \Phi _{i}$
contains the product $\varepsilon \beta _{ij}.$ That is, making the
expansion in powers of $\varepsilon ,$ Eq.\ (\ref{PExpansion}) within the
MFA results in the expressions that are also automatically expanded in $%
\beta _{ij}.$ In contrast, within the rigorous formalism parameters $%
\varepsilon $ and $\beta _{ij}$ are split from each other and the terms of
the $\varepsilon $ expansion contain general functions of $\beta _{ij}.$
Returning to the MFA, one obtains the expansion
\begin{eqnarray}
\tilde{a}_{i\uparrow }^{(1)}(t) &=&\int_{-\infty }^{t}dt^{\prime }e^{-i\Phi
_{i}^{(0)}(t^{\prime })}  \nonumber \\
\tilde{a}_{i\downarrow }^{(2)}(t) &=&\int_{-\infty }^{t}dt^{\prime }e^{i\Phi
_{i}^{(0)}(t^{\prime })}\tilde{a}_{i\uparrow }^{(1)}(t^{\prime })  \nonumber
\\
\tilde{a}_{i\uparrow }^{(3)}(t) &=&\int_{-\infty }^{t}dt^{\prime }e^{-i\Phi
_{i}^{(0)}(t^{\prime })}\left[ \tilde{a}_{i\downarrow }^{(2)}(t^{\prime
})\right.  \nonumber \\
&&+\left. \frac{4i}{\hbar }\sum_{j}J_{ij}\int_{0}^{t^{\prime }}dt^{\prime
\prime }\left| \tilde{a}_{i\uparrow }^{(1)}(t^{\prime \prime })\right| ^{2}%
\right] .  \label{aMFAExpansionRes}
\end{eqnarray}
Now the staying probability at the site $i$
\begin{equation}
P_{i}(t)=1-\left| \tilde{a}_{i\uparrow }(t)\right| ^{2}  \label{PiMFADef}
\end{equation}
can be expanded as
\begin{eqnarray}
P_{i}(t) &\cong &1-\left( \frac{\Delta }{2\hbar }\right) ^{2}\left| \tilde{a}%
_{i\uparrow }^{(1)}(t)\right| ^{2}  \nonumber \\
&&\quad +\left( \frac{\Delta }{2\hbar }\right) ^{4}2\func{Re}\left( \tilde{a}%
_{i\uparrow }^{(1)\ast }(t)\tilde{a}_{i\uparrow }^{(3)}(t)\right) ,
\label{PiMFAExpansion}
\end{eqnarray}
cf. Eq.\ (\ref{PiDelta4}). In the limit $t\rightarrow \infty ,$ the terms of
this formula that do not contain $J_{ij}$ just reproduce the known terms of
the expansion of the LZ probability $e^{-\varepsilon }$ in powers of $%
\varepsilon .$ One can rewrite it in the form
\begin{equation}
P_{i}\cong 1-\varepsilon +\frac{\varepsilon ^{2}}{2}+\varepsilon
^{2}\sum_{j}I_{ij},  \label{PiMFAepsExp}
\end{equation}
where
\begin{equation}
I_{ij}=\left( \frac{v}{2\pi \hbar }\right) ^{2}2\func{Re}\left( \tilde{a}%
_{i\uparrow }^{(1)\ast }(\infty )\tilde{b}_{ij\uparrow }^{(3)}(\infty
)\right)  \label{IijMFADef}
\end{equation}
and we have defined
\begin{equation}
\tilde{b}_{ij\uparrow }^{(3)}(\infty )\equiv \frac{4i}{\hbar }%
J_{ij}\int_{-\infty }^{\infty }dte^{-i\Phi
_{i}^{(0)}(t)}\int_{0}^{t}dt^{\prime }\left| \tilde{a}_{i\uparrow
}^{(1)}(t^{\prime })\right| ^{2}.  \label{bijMFADef}
\end{equation}

To calculate $I_{ij},$ it is convenient to introduce the dimensionless
parameters $\tilde{\alpha}_{i},$ $\beta _{ij},$ and $\tilde{\delta}_{ij}$
defined by Eq.\ (\ref{gammaijDef1}) and the dimensionless time variable
\begin{equation}
z\equiv \sqrt{\frac{v}{2\pi \hbar }}t.  \label{zTimeDef}
\end{equation}
Then Eq.\ (\ref{Phi0Def}) transforms to
\begin{equation}
\Phi _{i}^{(0)}(t)=\varphi _{i}(z)=\pi \left( z^{2}-2\tilde{\alpha}%
_{i}z\right) ,  \label{phiiMFADef}
\end{equation}
and from Eq.\ (\ref{IijMFADef}) one obtains
\begin{eqnarray}
I_{ij} &=&4\pi \beta _{ij}\func{Re}\left[ f_{i}^{(1)\ast }(\infty
)\int_{-\infty }^{\infty }dze^{-i\varphi _{i}(z)}\right.  \nonumber \\
&&\qquad \times \left. i\int_{0}^{z}dz^{\prime }\left| f_{j}^{(1)}(z^{\prime
})\right| ^{2}\right]  \label{AMFA}
\end{eqnarray}
and
\begin{equation}
f_{i}^{(1)}(z)\equiv \int_{-\infty }^{z}dz^{\prime }e^{-i\varphi
_{i}(z^{\prime })}.  \label{f1MFADef}
\end{equation}
One can integrate by parts in Eq.\ (\ref{AMFA}):
\begin{eqnarray}
I_{ij} &=&-4\pi \beta _{ij}\func{Re}\left[ if_{i}^{(1)\ast }(\infty )\right.
\nonumber \\
&&\qquad \times \left. \int_{-\infty }^{\infty }dzf_{i}^{(1)}(z)\left|
f_{j}^{(1)}(z)\right| ^{2}\right] .  \label{AMFA2}
\end{eqnarray}
Here $f_{i}^{(1)}(z)$ can be expressed in terms of Fresnel integrals $S(x)$
and $C(x)$:
\begin{eqnarray}
f_{i}^{(1)}(z) &=&\int_{-\infty }^{z}dz^{\prime }e^{-i\varphi _{i}(z^{\prime
})}=\int_{-\infty }^{z}dz^{\prime }e^{-i\pi \left( z^{\prime 2}-2\tilde{%
\alpha}_{i}z^{\prime }\right) }  \nonumber \\
&=&e^{i\pi \tilde{\alpha}_{i}^{2}}\int_{-\infty }^{z}dz^{\prime }e^{-i\pi
\left( z^{\prime }-\tilde{\alpha}_{i}\right) ^{2}}  \nonumber \\
&=&e^{i\pi \tilde{\alpha}_{i}^{2}}\int_{-\infty }^{z-\alpha _{i}}dz^{\prime
}e^{-i\pi z^{\prime 2}} \\
&=&\frac{e^{i\pi \tilde{\alpha}_{i}^{2}}}{\sqrt{2}}\left\{ \frac{1-i}{2}+C%
\left[ \sqrt{2}(z-\tilde{\alpha}_{i})\right] \right.  \nonumber \\
&&\qquad -\left. iS\left[ \sqrt{2}(z-\tilde{\alpha}_{i})\right] \right\} .
\label{f1Calc}
\end{eqnarray}
With $u=\sqrt{2}\left( z-\tilde{\alpha}_{j}\right) $ one obtains, finally
\begin{eqnarray}
I_{ij} &=&\frac{\pi \beta _{ij}}{\sqrt{2}}\int_{-\infty }^{\infty }du\left[
C(u-\sqrt{2}\tilde{\delta}_{ij})-S(u-\sqrt{2}\tilde{\delta}_{ij})\right]
\nonumber \\
&&\times \left[ \left( \frac{1}{2}+C(u)\right) ^{2}+\left( \frac{1}{2}%
+S(u)\right) ^{2}\right] .  \label{AijResShifted}
\end{eqnarray}
For $\tilde{\delta}_{ij}=0$ this result simplifies to
\begin{equation}
I_{ij}=\frac{\pi \beta _{ij}}{\sqrt{2}}\int_{-\infty }^{\infty }du\left[
C^{2}(u)-S^{2}(u)\right] =\beta _{ij}.  \label{AijRes1}
\end{equation}
Moreover, it can be shown that the symmetric part of Eq.\ (\ref
{AijResShifted}) defined by Eq.\ (\ref{IijSymDef}) coincides with the small-$%
\beta _{ij}$ expansion of the rigorous quantum mechanical result, Eqs.\ (\ref
{I0Def}) and (\ref{Iijweak}).

\bibliographystyle{prsty}
\bibliography{gar-oldworks,gar-books,gar-own,gar-spin,gar-tunneling,gar-lz,gar-relaxation}

\end{document}